\pdfoutput=1
\date{\today}

\documentclass[aps,prl,twocolumn,letterpaper,superscriptaddress,groupedaddress]{revtex4}
\usepackage{dcolumn,graphicx,amsmath,amssymb,algorithm,algpseudocode}
\usepackage{todonotes}
\usepackage{xcolor}
\usepackage[hidelinks]{hyperref}
\setcitestyle{super}

\newcommand{\ud}{\mathrm{d}}
\newcommand{\erf}{\mathrm{erf}}

\newcommand{\pdiff}[2]{\frac{\partial #1}{\partial #2}}

\begin{document}

\title{
Compensated Grid-Based Hybrid Exchange - 
A Simple, Accurate, Efficient, and Differentiable Recipe for Large-Scale
Molecular Systems
}

\author{Edward G. Hohenstein}
\author{Melisa Alkan}
\author{Xiaojie Wu}
\author{Robert M. Parrish}
\email[Corresponding author: ]{rparrish@nvidia.com}
\affiliation{Math Libraries, NVIDIA Corporation, 2788 San Tomas Expressway, Santa Clara, California 95051, USA}

\begin{abstract}
A compensated grid-based approach is developed for hybrid exchange in
large-scale molecular systems. The method uses a symmetrically
overlap-fitted grid-based exchange approach, with fitting performed in the
orthonormalized atomic orbital basis. This fitting is augmented with a
superposition-of-atomic-densities difference-density (dSCF) partitioning of the
exchange energy, yielding the finished compensated grid-based exchange approach. The
complete method is found to have orders of magnitude reduction of absolute and
relative error properties versus raw grid-based exchange, and remains
analytically differentiable. The method is easy to implement and provides
efficiency gains over GPU integral-direct techniques for high
quality and/or augmented basis sets. Library functions to help users implement
the method in their own quantum chemistry codes are provided in the
recently-shipped cuEST v0.3 library.  
\end{abstract}

\maketitle

\section{Introduction}

Hybrid exchange (historically denoted as ``K'') is a required component of a
modern density functional theory (DFT) toolkit. Since Becke's proposal
hybridizing pure DFT with Hartree--Fock exchange,\cite{Becke:1993:1372} yielding
the widely adopted B3PW91\cite{Becke:1993:5648} and
B3LYP\cite{Stephens:1994:11623} functionals, hybrid exchange has become
ubiquitous across dozens of popular density functionals. This has become doubly
important with the introduction of long-range-corrected (LRC) density
functionals,\cite{Savin:1996:327,Gill:1996:1005,Iikura:2001:3540} yielding the
workhorse $\omega$B97\cite{Chai:2008:084106} and
LC-$\omega$PBE\cite{Vydrov:2006:234109} families of density functionals, and
more lately typified by the extremely popular benchmark-grade $\omega$B97M-V
functional\cite{Mardirossian:2016:214110} and its COACH
extension.\cite{Liang:2026:2603.23466} Beyond resounding improvements
empirically observed in error properties,\cite{Goerigk:2017:32184} mixed hybrid
and LRC functionals have the compelling property that it is possible to recover
the asymptotically correct $100\%$ Hartree--Fock exchange limit at long range,
dramatically improving the stability properties of these functionals. This
manifests everywhere from improved treatment of charge-transfer excitations in
TD-DFT\cite{Dreuw:2003:2943,Tawada:2004:8425} to cleaned up ionization
properties in ground-state DFT to the fact that the SCF converges cleanly and
large organic systems remain insulators in everyday use of hybrid and LRC DFT.

That hybrid exchange is required does not imply that it is easy. Indeed, a
large body of work over the past few decades has been developed to mitigate the
computational cost of hybrid exchange on CPUs, and, more lately, on GPUs. For
smaller systems and/or larger basis sets, density fitting is an extremely
popular and successful
technique,\cite{Whitten:1973:4496,Dunlap:1979:3396,Vahtras:1993:514,Weigend:2002:4285,Polly:2004:2311}
with noteworthy CPU implementations in
TURBOMOLE,\cite{Balasubramani:2020:184107} Molpro,\cite{Werner:2020:144107}
Psi4,\cite{Smith:2020:184108} ORCA,\cite{Neese:2022:e1606}
Q-Chem,\cite{Epifanovsky:2021:084801} and PySCF,\cite{Sun:2020:024109} and GPU
implementations in GPU4PySCF,\cite{Wu:2025:e70008,Li:2025:1459}
TeraChem,\cite{Wang:2026:3398} and cuEST.\cite{cuEST} In the last of these
packages, additional gains are possible through the use of mixed precision Ozaki
tensor core emulation\cite{Ozaki:2012:95,Ozaki:2026:IJHPCA} accelerations of
the dense linear algebra steps, for which cuEST uses
cuBLAS,\cite{Schwarz:2026:91} an approach recently explored by other
authors.\cite{Huang:2026:3483} However, large RAM requirements for 3-index tensors
in the most popular flavors of density-fitted hybrid exchange preclude efficient
application to systems with more than a few hundred atoms. For larger systems,
integral-direct methods\cite{Almlof:1982:385,Haser:1989:104}
built on efficient integral algorithms\cite{McMurchie:1978:218,Rys:1983:154,Obara:1986:3963,Head-Gordon:1988:5777,Gill:1990:5564,Gill:1991:753}
are the standard choice. The integrals are re-computed exactly whenever they are required;
spatial sparsity\cite{Gill:1994:65,Schwegler:1996:2726,Ochsenfeld:1998:1663} and the 
fusion of integral evaluation with density contraction are used to accelerate integral-direct
exchange (and Coulomb) algorithms.
Integral-direct techniques are ubiquitous on CPUs, including in
Q-Chem,\cite{Epifanovsky:2021:084801} Gaussian,\cite{Gaussian16}
GAMESS,\cite{Barca:2020:154102} PySCF,\cite{Sun:2020:024109} and many other
codes. Integral-direct techniques were among the earliest targets of GPU quantum
chemistry, e.g., through the efforts of Mart\'{\i}nez and co-workers in
TeraChem\cite{Ufimtsev:2008:222,Ufimtsev:2009:1004} and many other authors in
codes like GAMESS,\cite{Asadchev:2010:696,Asadchev:2012:4166}
QUICK,\cite{Miao:2013:965,Manathunga:2020:4315}
EXESS,\cite{Barca:2020:SC20,Barca:2021:7486} Accelerated
DFT,\cite{Ju:2024:10838} and JoltQC.\cite{Wu:2026:194102} Integral-direct techniques are crucial for modern
quantum chemistry, and indeed the newest v0.3 version of cuEST (discussed within)
contains a new and efficient JIT-based library for integral-direct Coulomb and
exchange primitives. However, integral-direct approaches are often plagued by
reduced efficiency in high-angular-momentum and/or augmented basis sets, which are important
for the application of hybrid exchange with the most modern density functionals
like $\omega$B97M-V (requiring def2-TZVP or def2-QZVP quality basis sets for
accurate use).

A compelling alternative technique for hybrid exchange is to use a Becke-style
molecular quadrature grid for one of the two spatial integrals in the electron
repulsion integral tensor. An immediate advantage of this approach is that the
usual $\mathcal{O}(N^4)$ formal scaling of an integral-direct (or DF) exchange matrix
formation is reduced to $\mathcal{O}(N^3)$ -- before sparsity is considered.
The first incarnation of this idea is due to Friesner and is the pseudospectral approach
(PS)\cite{Friesner:1985:39,Friesner:1986:1462,Friesner:1987:3522} that forms
the basis of Schr\"odinger's Jaguar CPU code.\cite{Bochevarov:2013:2110} 
PS exchange as originally written goes considerably beyond simple grid-based
hybrid exchange in three respects.  First, one of the two collocation matrices
is transformed by a one-sided least-squares fit of the grid to the
non-normalized atomic orbitals.  Second, the fit is
dealiased:\cite{Friesner:1987:3522,Ringnalda:1990:3397} the least-squares fit is
performed in a basis augmented with dealiasing functions, whose contributions
are then explicitly removed, so that components of the orbital products outside
the atomic-orbital basis do not alias onto the fitted coefficients. Finally, PS
is noteworthy in that it typically uses varying amounts of 1-, 2-, and even
sometimes 3-center analytical corrections\cite{Greeley:1994:4028} whereby the
approximate PS electron repulsion integral tensor is partly replaced by
analytical corrections according to how many unique centers appear.  Analytical
corrections performed in this way are found to yield good error reduction
properties, but it can be difficult to form a variationally stable hybrid
exchange energy with these corrections, and it can be quite verbose to formulate
efficient codes for the needed energy and analytical gradient pieces with such
analytical corrections. The combination of fitting, dealiasing, and analytic
corrections defining Friesner's formulation of PS allows it to provide accurate
results with very coarse integration grids.

Interest in grid-based approaches to hybrid exchange was rejuvenated by
Neese and co-workers when they introduced the chain-of-spheres exchange (COSX)
method and implemented it in the ORCA CPU code.\cite{Neese:2009:98} Neese
removed much of the complexity of the earlier PS methods by using larger
integration grids and recovering the computational overhead through 
extensive exploitation of spatial sparsity in the grid-based integral
procedure.\cite{Neese:2009:98,Helmich-Paris:2021:104109} 
Later, COSX was extended to use a one-sided fitting of the grid to the atomic 
orbitals.\cite{Izsak:2011:144105} Without dealiasing functions, the fit can only 
correct components of the orbital products that lie within the span
of the orbital basis, so it primarily benefits larger basis sets;\cite{Holzer:2020:184115} 
complementary auxiliary basis sets have been considered as a means to span the missing 
space\cite{Izsak:2013:094111} and may be particularly useful in smaller basis sets.
Following the success of Neese's COSX method, a more widespread adoption of
these grid-based techniques -- a concept that is sometimes referred to as
``seminumerical exchange''\cite{Plessow:2012:810,Laqua:2018:3451} --
has emerged. The senex implementation in TURBOMOLE\cite{Plessow:2012:810}
emphasizes rigorous integral screening and the evaluation of local hybrid
functionals, as do related implementations for local hybrid
functionals.\cite{Bahmann:2015:1540,Liu:2017:2571} Also in TURBOMOLE,
Holzer's semiJK algorithm\cite{Holzer:2020:184115} treats both the Coulomb
and exchange matrices seminumerically for the many density matrices of
excited-state and property calculations, and introduces a symmetric,
overlap-metric de-aliasing of the collocation matrix.  Ochsenfeld and co-workers
developed the sn-LinK approach\cite{Laqua:2018:3451} which has been extended to
GPUs\cite{Laqua:2020:1456} and to mixed single/double
precision.\cite{Laqua:2021:214116} Very recently, a new CPU implementation of
sn-LinK has been developed in Q-Chem where it is known as
SNK.\cite{DiCenso:2026:2610.21075} Distributed-memory multi-GPU seminumerical
exchange has also been implemented in the GauXC
library.\cite{Williams-Young:2023:234104} Recently, a CPU implementation of
grid-based exchange based on sn-LinK\cite{Laqua:2018:3451} using the partition
scheme developed by Laqua {\em et al.}\cite{Laqua:2018:204111} and using the
advances described by Helmich-Paris {\em et al.}\cite{Helmich-Paris:2021:104109}
has been developed in PySCF\cite{Sun:2026:2603.14155} (known as SGX in that
package).  A combination of Neese's COSX with analytic corrections based on the
superposition of atomic densities (SAD) density matrix -- this CPU-based
approach is called aCOSx -- was developed by Liu and
co-workers;\cite{Zhang:2025:1492} this method was implemented in the Beijing
Density Functional (BDF) package.\cite{Zhang:2020:064113}
\textbf{\textit{Notational detail:}} Since the field has already settled on no
fewer than nine separate names for the concept of substituting a single Becke
grid into the ERIs in K, we have found it hard to pick a single one that
is more apt than the others. For several years, some of us have referred
to the raw concept of single Becke grid resolution without embellishments as
``grid-based hybrid exchange'' or ``Grid-K,'' a convention we use herein.

Another alternative worth a note is tensor hypercontraction
(THC),\cite{Hohenstein:2012:044103,Parrish:2012:224106,Hohenstein:2012:221101,Parrish:2013:132505}
which
conceptually resembles using Becke-style molecular quadrature grids in both
spatial integrals of the electron repulsion integral. Here one immediately
encounters the diagonal singularity of the Coulomb operator, and must
renormalize the grid representation of this operator through
least-squares\cite{Parrish:2012:224106,Hohenstein:2012:221101} or other
approaches.\cite{Lu:2015:329,Hu:2017:5420,Lee:2020:243} For more than a decade, THC has been a practical method for
post-Hartree--Fock
methods\cite{Hohenstein:2012:221101,Parrish:2014:181102,Hohenstein:2013:124111,Hohenstein:2013:12972,Song:2016:174111,Song:2017:034104,Song:2018:044108,Song:2020:234113}
where the high cost of the underlying method and smaller
magnitude of the correlation and/or excitation energy often yield efficient
performance versus accuracy characteristics.  However, the larger energy scales
and lower underlying scaling of SCF have contributed to the fact that the practical
application of THC to hybrid exchange has only recently shown significant
progress, through the least-norm regularized THC approach developed over the
last few years.\cite{Hillers-Bendtsen:2025:4734,Hillers-Bendtsen:2025:11595}
Note that the first THC-SCF paper used a flavor of the SAD difference density
dSCF procedure\cite{Parrish:2016:131101} employed herein to lower the magnitude of the quantity to be
estimated,\cite{Hillers-Bendtsen:2025:4734} but later refinement of the method
now takes one to four initial integral-direct SCF iterations to further lower
the precision requirements on the THC factorization.\cite{Hillers-Bendtsen:2025:11595}
THC-SCF is a promising direction for the field,
especially as only a handful of density fitting Coulomb metric analytical
integrals are needed for the THC part of the procedure, and as millihartree (m$E_{\rm h}$) accuracy for
THC-SCF is starting to become routinely possible. 

In the present work, we develop a compensated grid-based hybrid exchange
method, ``Compensated Grid-K,'' that combines two independent corrections to raw
grid-based exchange. The first is a SAD difference-density (dSCF) partition of
the exchange energy itself, rather than of the exchange matrix:\cite{Parrish:2016:131101} the exchange
energy of the SAD density matrix and its bilinear coupling
to the difference density are evaluated analytically, and only the small
quadratic term in the difference density is approximated on the grid. Because
the partition is applied to a well-defined energy functional, the resulting
exchange matrix is its exact derivative with respect to the density matrix and
the SCF procedure remains variational, while the analytic terms require nothing
beyond standard exchange-matrix builds for a fixed, position-independent
density matrix. Because the SAD density matrix is atom-block-diagonal, these 
analytic correction terms can be evaluated efficiently with any integral-direct
exchange algorithm that makes effective use of sparsity in the density matrix.
The second correction is a symmetric overlap fit of the grid
representation to the atomic orbitals, carried out in the canonically
orthogonalized basis. The fit is symmetric by construction, independent of the
ordering of the basis functions, and robust against the near-linear
dependencies that arise in large and augmented basis sets, which are precisely
the cases where integral-direct exchange is most expensive and where the fit
is most effective. Analytic nuclear gradients are derived for the complete
method, including the response of the overlap fit. Together, these corrections
reduce the absolute and relative errors of grid-based exchange by orders of
magnitude on coarse grids, and we demonstrate the accuracy of energies,
relative energies, and optimized geometries across standard benchmark sets in
double- through quadruple-zeta basis sets. All components are implemented on
NVIDIA GPUs and are provided as library functions in the cuEST v0.3 library.
We report timings for B3LYP energies and gradients on systems of more than
a thousand atoms. We believe the compensated Grid-K approach explored herein
has compelling accuracy versus cost properties, particularly in larger basis
sets with higher angular momentum functions, achieving near-microhartree
accuracy at a high speedup over GPU-based integral-direct algorithms.

The overlap fit used in compensated Grid-K invites comparison with
earlier fitting schemes for grid-based exchange, which differ in a few
essential choices. The fit may be global, with a single transfer matrix
spanning the full basis, or local, with separate fits over the basis functions
significant on each batch of grid points, which preserves the possibility of
linear scaling at the price of a fitted collocation matrix that depends on the
batching of the grid. The fit may be one-sided, dressing only the collocation
matrix that is not contracted with the density
matrix,\cite{Izsak:2011:144105,Izsak:2013:094111,Zhang:2025:1492} in which
case the permutational symmetry of the approximate ERIs must be restored
separately, or symmetric, dressing both collocation matrices identically.
Closest in spirit to the present approach is the elegant de-aliasing procedure
of Holzer's semiJK algorithm.\cite{Holzer:2020:184115} There, both collocation
matrices are dressed with the same matrix, obtained by solving a triangular
system between the Cholesky factors of the analytic and quadrature overlap
matrices, so that the fitted quadrature overlap reproduces the analytic
overlap exactly and the permutational symmetry of the approximate ERIs is
retained by construction. The procedure requires nothing beyond the orbital
basis set and the quadrature grid, is applied locally over the basis functions
significant on each batch of grid points, and extends naturally to the many
density matrices of excited-state and property calculations, for which Holzer
demonstrated accuracy competitive with density fitting. Holzer also made the
important observation that the correction is most effective in polarized
triple- and quadruple-zeta basis sets, where the orbital basis provides a
sufficiently rich expansion space, and offers less in smaller basis sets -- a
trend we also observe for compensated Grid-K below. The transfer matrix used
in compensated Grid-K satisfies the same symmetric constraint but is formulated 
differently: it is the symmetric inverse square root of the quadrature metric in 
the canonically orthogonalized basis, which makes it independent of the ordering
of the basis functions and robust to near-linear dependence through the
orthogonalization threshold, and it is global, so that its response to the
atomic positions can be written in closed form for the analytic gradient.

We also note similarities between this approach and the aCOSx method of Liu and
co-workers.\cite{Zhang:2025:1492} There, the density matrix is also
partitioned into a SAD density matrix and a remainder term.
The exchange matrix of the SAD part is evaluated analytically at the start
of the SCF procedure and only the exchange matrix of the remainder term is approximated by COSX.
Similar to the compensated Grid-K approach presented here, the remainder is small 
and tolerates a coarser approximation. The differences, however, are more
fundamental than the similarities. Liu and co-workers partition the exchange
matrix rather than the exchange energy: no approximate energy is ever defined,
the bilinear term between the atomic and remainder density matrices is
unspecified, and contracting the partitioned exchange matrix with the density
does not yield an energy of which that matrix is the derivative. In dSCF the
energy is defined first, the bilinear term is exact, and the exchange matrix
and gradient follow from it.\cite{Parrish:2016:131101} The aCOSx energy is
moreover not variational: after convergence, the remainder exchange is
re-evaluated post hoc, analytically for near atomic orbital pairs, so the
reported energy belongs to a different approximation than the one for which
the orbitals are stationary, and a rigorous analytic gradient would require
coupled-perturbed SCF response.

\section{Theory}

\subsection{Hybrid Exchange}

We define the $3 N_{\mathrm{atom}}$ atomic coordinates $\{ \vec
r_{\mathcal{A}} \} \equiv \{ r_{\mathcal{A}, i} \} \equiv \{ r_\xi \}$ (note
that here and throughout $\equiv$ is read as ``defined to be,'' which some
authors denote as $:=$).

For an atom-centered basis set of real contracted Gaussians $\{ \phi_{p} (\vec
r_1; \{ r_\xi \}) \}$ (the ``spatial atomic orbital basis set''), the electron
repulsion integral tensor is,
\begin{equation}
\label{eq:eri}
(pq|\hat o|rs)
\equiv
\iint_{\mathbb{R}^6}
\ud \vec r_1 \
\ud \vec r_2 \
\phi_{p} (\vec r_1)
\phi_{q} (\vec r_1)
\hat o (r_{12})
\phi_{r} (\vec r_2)
\phi_{s} (\vec r_2)
\end{equation}
Here, the generalized Coulomb interaction kernel is,
\begin{equation}
\label{eq:lrc-kernel}
\hat o(
r_{12} ;
\alpha,
\beta,
\omega)
\equiv
\alpha
\frac{1}{r_{12}}
+
\beta
\frac{\erf(\omega r_{12})}{r_{12}}
\end{equation}
Setting $\alpha \equiv 1, \beta \equiv 0$ yields the usual Coulomb operator,
while other choices allow for the full range of common hybrid
long-range-corrected (LRC) exchange operators. Note that we consider only
$\alpha \geq 0$, $\alpha + \beta \geq 0$, $\omega \geq 0$ throughout for a
sensible positive semidefinite generalized Coulomb operator. With this
generalized definition of the Coulomb interaction, we will refer to any mix of
Hartree--Fock and/or LRC exchange as simply ``hybrid exchange'' hereafter.

For a one-particle density matrix $D_{pq}$ (here and throughout, illustrated in
the alpha spin channel), the hybrid exchange energy is,
\begin{equation}
\label{eq:hybrid-exchange-energy}
E_{K} [D_{pq}]
\equiv
\frac{1}{2}
\sum_{pq}
\sum_{rs}
D_{pq}
(pr|\hat o|qs)
D_{rs}
\end{equation}
The hybrid exchange matrix (i.e., the hybrid exchange contribution to the Fock
matrix) is,
\begin{equation}
\label{eq:hybrid-exchange-matrix}
K_{pq} [D_{rs}]
\equiv
\pdiff{
E_{K}
}{
D_{pq}
}
=
\sum_{rs}
(pr|\hat o|qs)
D_{rs}
\end{equation}
The hybrid exchange atomic gradient is,
\begin{equation}
\nonumber
E_{K}^{\xi} [D_{pq}]
\equiv
\pdiff{
E_{K}
}{
r_\xi
}
=
\frac{1}{2}
\sum_{pq}
\sum_{rs}
D_{pq}
\pdiff{
(pr|\hat o|qs)
}{
r_\xi
}
D_{rs}
\end{equation}
\begin{equation}
\label{eq:hybrid-exchange-nuclear-gradient}
=
2
\sum_{pq}
\sum_{rs}
D_{pq}
(\left [\partial_{r_\xi} p\right ]r|\hat o|qs)
D_{rs}
\end{equation}
Note that in Eqs.~\eqref{eq:hybrid-exchange-matrix} and
\eqref{eq:hybrid-exchange-nuclear-gradient}, the primary definition is of the
derivative of the exchange energy with respect to a density matrix or atomic
coordinate perturbation. E.g., the last expression in the atomic gradient with
four symmetric contributions from the basis function advection is a consequence
of our current definition of the exchange energy and atomic orbital basis set.
When we later approximate or re-partition the exchange energy, these derivatives
must be taken from scratch to yield a consistent theory. 

Note that, for the moment,
\begin{equation}
\label{eq:EKdot}
E_{K}
=
\frac{1}{2}
\sum_{pq}
D_{pq}
K_{pq}
\end{equation}
This identity follows from $E_{K}$ being a homogeneous quadratic form in
$D_{pq}$. It no longer holds once the density matrix is partitioned and the ERIs
are approximated, as in the dSCF formulation below.

Note also that the hybrid exchange energy enters the density functional theory
total energy expression with a negative sign (i.e., the exchange contribution to
the total energy is negative definite). This is omitted here and throughout for
brevity. 

\subsection{Superposition of Atomic Densities Difference Density Picture (dSCF)}

Before approximating the ERIs, it is worth asking if such approximations could be
applied to a smaller quantity than $E_{K}$, while preserving the
differentiability and computational efficiency of the overall procedure? For the
common case of hybrid density functional theory where $D_{pq}$ is the full
alpha-spin one particle density matrix of all alpha electrons in the system, the
answer is yes, but the approach is slightly subtle. 

As was first pointed out by Becke in the context of the NUMOL Coulomb (J)
solver,\cite{Becke:1988:2993} and later formalized by some of us as the
``difference density self-consistent field picture,''
(dSCF)\cite{Parrish:2016:131101} the one-particle density matrix of
density functional theory can be advantageously partitioned into a superposition
of atomic densities (SAD) component $D_{pq}^{\varnothing}$ and a difference
component,
$D_{pq}^{\Delta}$,
\begin{equation}
D_{pq}
\equiv
D_{pq}^{\varnothing}
+
D_{pq}^{\Delta}
\end{equation}
Here $D_{pq}^{\varnothing}$ is an atom-block-diagonal ($p$ and $q$ are on the
same atomic center for all nonzero blocks) density matrix of
spherically-isotropic, fractionally-occupied orbitals chosen to yield
a neutral atom. An important property of $D_{pq}^{\varnothing}$ is that it is
invariant to the atomic positions, and therefore has no response contributions
in derivative properties. $D_{pq}^{\varnothing}$ can be tabulated for a given
basis set or chosen by one of many possible heuristic procedures. One approach
that we find works well is Knizia's suggestion\cite{Knizia:2013:4834} to
compute the corresponding orbitals of the Aufbau-occupied natural orbitals of a
reference basis (e.g., the fully contracted form of cc-pVDZ\cite{Dunning:1989}
or ANO-DK3\cite{Tsuchiya:2001:4463}) within the primary basis at hand. Note
that this approach is widely used across Gaussian-basis quantum chemistry to
form the superposition of atomic densities guess for the first iteration of the
density functional theory procedure. 

With $D_{pq}^{\varnothing}$ defined, $D_{pq}^{\Delta}$ is exactly determined by
the additivity constraint. Physically $D_{pq}^{\Delta}$ represents the
polarization of the density matrix from the SAD density matrix, including any
net electron attachment or removal for the case of cations, anions, and/or
net-high-spin radicals (note that one could attempt to rescale the SAD density
to account for non-neutral total states, but we find that this is unnecessary
and overly complicated). 

In dSCF, the crucial trick is that prior to ERI approximation, one must start
again from the energy expression,
\begin{equation}
\nonumber
E_{K}^{\mathrm{approx}} [D_{pq}^{\Delta} ; D_{pq}^{\varnothing}]
\stackrel{\approx}{\equiv}
\frac{1}{2}
\sum_{pq}
\sum_{rs}
D_{pq}^{\varnothing}
(pr|\hat o|qs)
D_{rs}^{\varnothing}
\end{equation}
\begin{equation}
\nonumber
+
\sum_{pq}
\sum_{rs}
D_{pq}^{\varnothing}
(pr|\hat o|qs)
D_{rs}^{\Delta}
\end{equation}
\begin{equation}
+
\frac{1}{2}
\sum_{pq}
\sum_{rs}
D_{pq}^{\Delta}
(pr|\hat o|qs)^{\mathrm{approx}}
D_{rs}^{\Delta}
\end{equation}
Here the symbol $\stackrel{\approx}{\equiv}$ means ``is defined to be
approximated as,'' with all derivative quantities following exactly after the
approximated quantity is defined.  
Now the dSCF hybrid exchange matrix is,
\begin{equation}
\nonumber
K_{pq}^{\mathrm{approx}}
[D_{rs}^{\Delta} ; D_{rs}^{\varnothing}]
\equiv
\pdiff{
E_{K}^{\mathrm{approx}}
}{
D_{pq}^{\Delta}
}
\end{equation}
\begin{equation}
\nonumber
=
\sum_{rs}
(pr|\hat o|qs)
D_{rs}^{\varnothing}
+
\sum_{rs}
(pr|\hat o|qs)^{\mathrm{approx}}
D_{rs}^{\Delta}
\end{equation}
\begin{equation}
\nonumber
=
K_{pq}
[ D_{rs}^{\varnothing} ]
+
K_{pq}^{\mathrm{approx}}
[ D_{rs}^{\Delta} ]
\end{equation}
\begin{equation}
\equiv
K_{pq}^{\varnothing}
[ D_{rs}^{\varnothing} ]
+
K_{pq}^{\Delta, \mathrm{approx}}
[ D_{rs}^{\Delta} ]
\end{equation}
The dSCF hybrid exchange atomic gradient is,
\begin{equation}
\nonumber
E_{K}^{\mathrm{approx}, \xi} [D_{pq}^{\Delta} ; D_{pq}^{\varnothing}]
\equiv
\pdiff{
E_{K}^{\mathrm{approx}}
}{
r_{\xi}
}
\end{equation}
\begin{equation}
\nonumber
=
\frac{1}{2}
\sum_{pq}
\sum_{rs}
D_{pq}^{\varnothing}
\pdiff{
(pr|\hat o|qs)
}{
r_\xi
}
D_{rs}^{\varnothing}
\end{equation}
\begin{equation}
\nonumber
+
\sum_{pq}
\sum_{rs}
D_{pq}^{\varnothing}
\pdiff{
(pr|\hat o|qs)
}{
r_\xi
}
D_{rs}^{\Delta}
\end{equation}
\begin{equation}
+
\frac{1}{2}
\sum_{pq}
\sum_{rs}
D_{pq}^{\Delta}
\pdiff{
(pr|\hat o|qs)^{\mathrm{approx}}
}{
r_\xi
}
D_{rs}^{\Delta}
\end{equation}

There are a few key features of this approach worth noting: 
\begin{enumerate}
\item Now only $E_{K}^{\Delta} \equiv \frac{1}{2} \sum_{pq} \sum_{rs}
D_{pq}^{\Delta} (pr|\hat o|qs) D_{rs}^{\Delta}$ is approximated - this quantity
is generally orders of magnitude smaller than $E_{K}$ and so tolerates coarser
relative approximation.
\item $E_{K}^{\Delta}$ is quadratic in
$D_{pq}^{\Delta}$, just as $E_{K}$ was in $D_{pq}$.
\item If $(pr|\hat o|qs)^{\mathrm{approx}}$ is differentiable, so too is
$E_{K}^{\mathrm{approx}}$. Any orbital response contributions are as they were
due to the linearity $D_{pq} = D_{pq}^{\varnothing} + D_{pq}^{\Delta}$.
For standard flavors of hybrid Kohn--Sham density functional theory, such orbital
response contributions are all encoded in the energy weighted density matrix
overlap integral gradient contributions as before.
\item Crucially, $E_{K}^{\mathrm{approx}} \neq \frac{1}{2} \sum_{pq} D_{pq}
K_{pq}^{\mathrm{approx}} [D_{rs}^{\Delta} ; D_{rs}^{\varnothing}]$! This
formula, which is the doppelg\"anger of the linear dot product identity of
Eq.~\eqref{eq:EKdot} in the usual hybrid exchange, does not yield the
variational dSCF hybrid exchange energy due to the way it treats the bilinear
terms (one approximated, one exact in the dot product formula, both exact in
dSCF). This is an extremely important consideration - dSCF is not an
approximation to the K matrix, but a reformulation of the SCF hybrid exchange
energy formula that yields a precise, variational, and differentiable electronic
response surface.   
\end{enumerate}

\subsection{Grid-Based Exchange}

A very popular idea is to approximate one of the two spatial integrals in the
ERI with a Becke-style molecular quadrature grid\cite{Becke:1988:2547} $\{ <\vec r_A, w_A > \}$,
noting that the grid points and weights both depend on the atomic
positions $\{ r_{\xi} \}$,
\begin{equation}
(pq|\hat o|rs)^{\mathrm{grid}}
\equiv
P_{pq|rs}
\sum_{A}
w_{A}
X_{p}^{A}
X_{q}^{A}
V_{rs}^{A}
\end{equation}
The permutation operator is $P_{pq|rs} T_{pq|rs} \equiv \frac{1}{2} (T_{pq|rs} +
T_{rs|pq})$.
Here and throughout, the potential integrals are,
\begin{equation}
V_{rs}^{A}
\equiv
\int_{\mathbb{R}^3}
\ud \vec r_2 \
\hat o (r_{A2})
\phi_{r} (\vec r_2)
\phi_{s} (\vec r_2)
\end{equation}
Without any further modification, the collocation matrix is,
\begin{equation}
\label{eq:collocation}
X_{p}^{A}
\equiv
\phi_{p} (\vec r_A)
\end{equation}
The corresponding Grid-K approximation to the hybrid exchange energy is,
\begin{equation}
\label{eq:gridk_E}
E_{K}^{\mathrm{grid}}
[ D_{pq} ]
\stackrel{\approx}{\equiv}
\frac{1}{2}
\sum_{pq}
\sum_{rs}
D_{pq}
(pr|\hat o|qs)^{\mathrm{grid}}
D_{rs}
\end{equation}
The collocation matrices carry the two basis
functions of the electron integrated by quadrature and the potential integrals
those of the electron integrated analytically. Contracting the density matrix
into the collocation matrix and then into the potential integrals gives the
density-contracted collocation matrix and the exchange potential on the grid,
\begin{equation}
\label{eq:gridk_Y}
Y_{q}^{A}
\equiv
\sum_{p}
X_{p}^{A}
D_{pq}
\end{equation}
\begin{equation}
\label{eq:gridk_Z}
Z_{q}^{A}
\equiv
\sum_{s}
V_{qs}^{A}
Y_{s}^{A}
\end{equation}
The Grid-K exchange matrix is,
\begin{equation}
\label{eq:gridk_K}
K_{pq}^{\mathrm{grid}}
[ D_{rs} ]
\equiv
\pdiff{
E_{K}^{\mathrm{grid}}
}{
D_{pq}
}
=
\frac{1}{2}
\sum_{A}
w_{A}
\left [
X_{p}^{A}
Z_{q}^{A}
+
X_{q}^{A}
Z_{p}^{A}
\right ]
\end{equation}
And the exchange energy dot product of Eq.~\eqref{eq:EKdot} survives the grid
approximation,
\begin{equation}
\label{eq:gridk_Edot}
E_{K}^{\mathrm{grid}}
[ D_{pq} ]
=
\frac{1}{2}
\sum_{pq}
D_{pq}
K_{pq}^{\mathrm{grid}}
\end{equation}

\subsection{Symmetrically Overlap-Fitted Grid-Based Exchange}

As written above, the grid-based exchange approach will converge to the exact
result as the quadrature grid quality is increased. However, it is prudent to
ask if we might achieve better accuracy for a given intermediate-sized
quadrature grid at the same or similar computational cost by tailoring the grid
to the basis functions at hand. One popular trick along these lines is to adjust
the $\hat X$ matrix to exactly fit one or more analytically-known targets, such
as the overlap matrix,
\begin{equation}
S_{pq}
\equiv
\int_{\mathbb{R}^3}
\ud \vec r_1 \
\phi_{p} (\vec r_1)
\phi_{q} (\vec r_1)
\approx
S_{pq}^{\mathrm{grid}}
\equiv
\sum_{A}
w_{A}
X_{p}^{A}
X_{q}^{A}
\end{equation}
There are many ways to justify this heuristic adjustment of the collocation
matrix to satisfy the overlap matrix integration constraint -- a very popular
one is that this ensures that the quadrature grid is adjusted to exactly
integrate the correct number of electrons for any density matrix.

Typically this is done asymmetrically, e.g., by replacing the two instances of
$\hat X$ in the Grid-K formula with separate instances of $\bar Q$ and $\bar R$,
\begin{equation}
S_{pq}
=
\sum_{A}
w_{A}
\bar Q_{p}^{A}
\bar R_{q}^{A}
\end{equation}
Indeed, one most often encounters this with $R_{q}^{A} \equiv X_{q}^{A}$ to
preserve the original collocation matrix contracted through the density matrix
in the formation of the exchange matrix.  Moreover, the fitting to determine
$\bar Q$ and $\bar R$ is typically done in the raw AO basis, which is very ill
conditioned. The asymmetrical fitting reduces the symmetry of the approximate ERIs
further and it is correctly restored by averaging the permutations of the $\bar{Q}$
and $\bar{R}$ matrices -- this roughly doubles the computational cost for symmetric density
matrices. Avoiding the extra computation by symmetrizing the resultant exchange
matrix yields the same exchange energy but an exchange matrix that is not its
derivative with respect to the density matrix, so the SCF is no longer
variational.
We note this asymmetrical AO-basis fitting procedure to provide
historical context, but do not refer to this approach throughout the rest of the
text.

We choose instead to use the symmetrical overlap-fitting formula,
\begin{equation}
S_{pq}
=
\sum_{A}
w_{A}
\bar X_{p}^{A}
\bar X_{q}^{A}
\end{equation}
and moreover choose the fitting procedure to match the grid to the canonically
orthogonalized atomic orbitals,
\begin{equation}
\label{eq:Xbar}
\bar X_{p}^{A}
\equiv
\sum_{q}
M_{pq}
X_{q}^{A}
\end{equation}
Here the transfer matrix is,
\begin{equation}
\label{eq:M}
M_{pq}
\equiv
\sum_{p'q'}
U_{pp'}
s_{p'}^{+1/2}
\Omega_{p'q'}^{-1/2}
s_{q'}^{-1/2}
U_{qq'}
\end{equation}
The eigendecomposition of the analytical overlap matrix is,
\begin{equation}
S_{pq}
\equiv
\sum_{r'}
U_{pr'}
s_{r'}
U_{qr'}
\end{equation}
And only eigenvalues greater than $\tau$ (the canonical atomic
orbital orthogonalization threshold) are retained. The matrix elements of the
inverse square root of the metric are read as $\Omega_{p'q'}^{-1/2} \equiv [\hat
\Omega^{-1/2}]_{p'q'}$. The pseudo-positive definite metric is,
\begin{equation}
\label{eq:Omega}
\Omega_{p'q'}
\equiv
\sum_{pq}
s_{p'}^{-1/2}
U_{pp'}
S_{pq}^{\mathrm{grid}}
U_{qq'}
s_{q'}^{-1/2}
\approx
\delta_{p'q'}
\end{equation}
Here ``pseudo-'' refers to the fact that negative quadrature weights are sometimes
encountered in Becke quadrature grids, usually due to the Lebedev angular
quadrature rules (e.g., the 74, 230, 266 point rules). These are always
encountered in concert with many additional positive quadrature weights in the
same Lebedev grid so that any sensible inner product evaluated with this
quadrature grid should yield a numerically positive definite metric matrix.
If negative eigenvalues are observed in $\hat \Omega$, the underlying quadrature
grid is likely very far from the needed quality to perform the fitting, and the
code should return an error. 

It is important to use the same set of canonical orthogonal orbitals within the
overlap-fitted Grid-K procedure and throughout the quantum chemistry procedure for
cases where linear dependencies are removed by the $\tau$ threshold in the
canonical orthogonalization step. Ideally any code implementation diagonalizes
$S_{pq}$ once and uses a single eigendecomposition and selected canonical
atomic orbitals for all steps. 

The transfer matrix enters the formation of the exchange matrix only through
the collocation matrix. Replacing $X_{p}^{A}$ with the dressed $\bar X_{p}^{A}$
in the grid ERIs and in Eqs.~\eqref{eq:gridk_E}--\eqref{eq:gridk_Edot} gives
the modified Grid-K ERIs, exchange energy, intermediates, and exchange matrix,
\begin{equation}
(pq \bar{|} \hat o \bar{|} rs)^{\mathrm{grid}}
\equiv
P_{pq|rs}
\sum_{A}
w_{A}
\bar X_{p}^{A}
\bar X_{q}^{A}
V_{rs}^{A}
\end{equation}
\begin{equation}
\label{eq:cgridk_E}
\bar E_{K}^{\mathrm{grid}}
[ D_{pq} ]
\stackrel{\approx}{\equiv}
\frac{1}{2}
\sum_{pq}
\sum_{rs}
D_{pq}
(pr \bar{|} \hat o \bar{|} qs)^{\mathrm{grid}}
D_{rs}
\end{equation}
\begin{equation}
\label{eq:cgridk_Y}
\bar Y_{q}^{A}
\equiv
\sum_{p}
\bar X_{p}^{A}
D_{pq}
\end{equation}
\begin{equation}
\label{eq:cgridk_Z}
\bar Z_{q}^{A}
\equiv
\sum_{s}
V_{qs}^{A}
\bar Y_{s}^{A}
\end{equation}
\begin{equation}
\bar K_{pq}^{\mathrm{grid}}
[ D_{rs} ]
\equiv
\frac{1}{2}
\sum_{A}
w_{A}
\left [
\bar X_{p}^{A}
\bar Z_{q}^{A}
+
\bar X_{q}^{A}
\bar Z_{p}^{A}
\right ]
\end{equation}
\begin{equation}
\bar E_{K}^{\mathrm{grid}}
[ D_{pq} ]
=
\frac{1}{2}
\sum_{pq}
D_{pq}
\bar K_{pq}^{\mathrm{grid}}
\end{equation}
The quadrature weights, potential integrals, and contraction pattern are
untouched, so an existing Grid-K exchange energy and exchange matrix code is
modified only by dressing its collocation matrix with $M_{pq}$.

\subsection{Grid-Based Exchange Gradient}

The purpose of this section is to show how to evaluate the analytical atomic
gradient of the Grid-K approximated exchange energy, obtained by differentiating
the approximate ERIs in Eq.~\eqref{eq:gridk_E} at fixed density matrix,
\begin{equation}
\nonumber
E_{K}^{\mathrm{grid}, \xi}
[ D_{pq} ]
\equiv
\pdiff{
E_{K}^{\mathrm{grid}}
}{
r_\xi
}
=
\frac{1}{2}
\sum_{pq}
\sum_{rs}
D_{pq}
\end{equation}
\begin{equation}
\times
\left [
\pdiff{}{r_\xi}
P_{pr|qs}
\sum_{A}
w_{A}
X_{p}^{A}
X_{r}^{A}
V_{qs}^{A}
\right ]
D_{rs}
\end{equation}
The Grid-K exchange energy depends on the atomic positions through the
quadrature weights (particularly in the Becke Voronoi weights, including both
grid point advection and Voronoi center advection terms), the collocation matrix
(including both basis function advection and grid point advection terms), and
the potential integrals (including both the basis function advection and grid
point advection terms), and it can be written as a contraction of each of these
against a corresponding pseudo-density,
\begin{equation}
\nonumber
E_{K}^{\mathrm{grid}}
[ D_{pq} ]
=
\sum_{A}
\epsilon_{A}
w_{A}
\end{equation}
\begin{equation}
\nonumber
=
\frac{1}{2}
\sum_{A}
\sum_{p}
X_{p}^{A}
\Lambda_{p}^{A}
\end{equation}
\begin{equation}
\label{eq:gridk_Edecomp}
=
\sum_{A}
\sum_{pq}
V_{pq}^{A}
\Gamma_{pq}^{A}
\end{equation}
Here the exchange pseudo-energy density, the collocation pseudo-density, and
the factored potential pseudo-density are,
\begin{equation}
\label{eq:eps}
\epsilon_{A}
\equiv
\frac{1}{2}
\sum_{q}
Y_{q}^{A}
Z_{q}^{A}
\end{equation}
\begin{equation}
\label{eq:Lambda}
\Lambda_{p}^{A}
\equiv
w_{A}
\sum_{q}
D_{pq}
Z_{q}^{A}
\end{equation}
\begin{equation}
\label{eq:Gamma}
\Gamma_{pq}^{A}
\equiv
\frac{1}{2}
w_{A}
Y_{p}^{A}
Y_{q}^{A}
\end{equation}
Each pseudo-density is the derivative of the Grid-K exchange energy with
respect to the quantity it is contracted against. The factor of one half in
the collocation contraction reflects that the energy is quadratic in the
collocation matrix and linear in the weights and potential integrals. The
Grid-K exchange gradient is therefore the same three contractions, with the
quadrature weights, collocation matrix, and potential integrals replaced by
their derivatives with respect to the atomic coordinates,
\begin{equation}
\nonumber
E_{K}^{\mathrm{grid}, \xi}
[ D_{pq} ]
=
\sum_{A}
\epsilon_{A}
\pdiff{w_{A}}{r_\xi}
+
\sum_{A}
\sum_{p}
\Lambda_{p}^{A}
\pdiff{X_{p}^{A}}{r_\xi}
\end{equation}
\begin{equation}
\label{eq:gridk_grad}
+
\sum_{A}
\sum_{pq}
\Gamma_{pq}^{A}
\pdiff{V_{pq}^{A}}{r_\xi}
\end{equation}
The factor of one half on the collocation term is absent because both
collocation matrices in the energy are differentiated.

Note that it is very tempting to assume that the Becke molecular quadrature grid
is complete and to retain only basis function advection terms, and even to
possibly go beyond this and treat basis function advection term contributions as
being identical from the collocation and potential integral parts, retaining
only the collocation advections for computational simplicity. This is the
approach taken in a recent GPU implementation of seminumerical exchange
gradients,\cite{Mironov:2026:tt68b} where only the collocation-side basis
function derivatives are retained and no derivatives of the potential integrals
or of the grid are formed. However, such a heuristic is not an analytical
gradient in the limit of any finite Becke molecular quadrature. 

\subsection{Symmetrically Overlap-Fitted Grid-Based Exchange Gradient}

The only difference from the previous section is that the dressed
collocation matrix of Eq.~\eqref{eq:Xbar} depends on the atomic positions
through both the raw collocation matrix and the transfer matrix.
Differentiating Eq.~\eqref{eq:cgridk_E} at fixed density matrix,
\begin{equation}
\nonumber
\bar E_{K}^{\mathrm{grid}, \xi}
[ D_{pq} ]
\equiv
\pdiff{
\bar E_{K}^{\mathrm{grid}}
}{
r_\xi
}
=
\frac{1}{2}
\sum_{pq}
\sum_{rs}
D_{pq}
\end{equation}
\begin{equation}
\times
\left [
\pdiff{}{r_\xi}
P_{pr|qs}
\sum_{A}
w_{A}
\bar X_{p}^{A}
\bar X_{r}^{A}
V_{qs}^{A}
\right ]
D_{rs}
\end{equation}
The modified energy therefore has a fourth ingredient, the transfer matrix,
with its own pseudo-density,
\begin{equation}
\nonumber
\bar E_{K}^{\mathrm{grid}}
[ D_{pq} ]
=
\sum_{A}
\bar \epsilon_{A}
w_{A}
\end{equation}
\begin{equation}
\nonumber
=
\frac{1}{2}
\sum_{A}
\sum_{p}
X_{p}^{A}
\Lambda_{p}^{A}
\end{equation}
\begin{equation}
\nonumber
=
\sum_{A}
\sum_{pq}
V_{pq}^{A}
\bar \Gamma_{pq}^{A}
\end{equation}
\begin{equation}
\label{eq:cgridk_Edecomp}
=
\frac{1}{2}
\sum_{pq}
M_{pq}
\Pi_{pq}
\end{equation}
The pseudo-energy density and the factored potential pseudo-density are
those of Eqs.~\eqref{eq:eps} and \eqref{eq:Gamma} with the dressed
intermediates of Eqs.~\eqref{eq:cgridk_Y} and \eqref{eq:cgridk_Z}, and the
same replacement in Eq.~\eqref{eq:Lambda} gives the pseudo-density of the
dressed collocation matrix,
\begin{equation}
\bar \epsilon_{A}
\equiv
\frac{1}{2}
\sum_{q}
\bar Y_{q}^{A}
\bar Z_{q}^{A}
\end{equation}
\begin{equation}
\bar \Lambda_{p}^{A}
\equiv
w_{A}
\sum_{q}
D_{pq}
\bar Z_{q}^{A}
\end{equation}
\begin{equation}
\bar \Gamma_{pq}^{A}
\equiv
\frac{1}{2}
w_{A}
\bar Y_{p}^{A}
\bar Y_{q}^{A}
\end{equation}
Because the dressed collocation matrix is linear in both the raw collocation
matrix and the transfer matrix, the energy is quadratic in each, hence the
two factors of one half above. The pseudo-density of the dressed collocation
matrix is back-transformed by the transfer matrix to give that of the raw
collocation matrix, and contracted against the raw collocation matrix to give
the transfer matrix pseudo-density,
\begin{equation}
\Lambda_{p}^{A}
\equiv
\sum_{q}
M_{qp}
\bar \Lambda_{q}^{A}
\end{equation}
\begin{equation}
\label{eq:Pi}
\Pi_{pq}
\equiv
\sum_{A}
\bar \Lambda_{p}^{A}
X_{q}^{A}
\end{equation}
The gradient is the three contractions of Eq.~\eqref{eq:gridk_grad}, with
the pseudo-densities above, plus the response of the transfer matrix,
\begin{equation}
\nonumber
\bar E_{K}^{\mathrm{grid}, \xi}
[ D_{pq} ]
=
\sum_{A}
\bar \epsilon_{A}
\pdiff{w_{A}}{r_\xi}
+
\sum_{A}
\sum_{p}
\Lambda_{p}^{A}
\pdiff{X_{p}^{A}}{r_\xi}
\end{equation}
\begin{equation}
\label{eq:cgridk_grad}
+
\sum_{A}
\sum_{pq}
\bar \Gamma_{pq}^{A}
\pdiff{V_{pq}^{A}}{r_\xi}
+
\sum_{pq}
\Pi_{pq}
\pdiff{M_{pq}}{r_\xi}
\end{equation}

We now derive an expression for the last term of Eq.~\eqref{eq:cgridk_grad},
starting from the energy in terms of the transfer matrix and its
pseudo-density, the last line of Eq.~\eqref{eq:cgridk_Edecomp}. By
Eqs.~\eqref{eq:M} and \eqref{eq:Omega}, the transfer matrix depends on the
atomic positions through the raw collocation matrix and the quadrature
weights, which enter only through the metric, and now also through the
overlap matrix, which enters both through the metric and directly through
its eigenvectors and eigenvalues. The last term of
Eq.~\eqref{eq:cgridk_grad} is therefore a sum of three contractions,
\begin{equation}
\nonumber
\sum_{pq}
\Pi_{pq}
\pdiff{M_{pq}}{r_\xi}
=
\sum_{A}
\tilde \epsilon_{A}
\pdiff{w_{A}}{r_\xi}
+
\sum_{A}
\sum_{p}
\tilde \Lambda_{p}^{A}
\pdiff{X_{p}^{A}}{r_\xi}
\end{equation}
\begin{equation}
\label{eq:dM_decomp}
+
\sum_{pq}
\Sigma_{pq}
\pdiff{S_{pq}}{r_\xi}
\end{equation}
which add a transfer matrix contribution to the pseudo-energy density and to
the collocation pseudo-density and introduce an overlap pseudo-density. All
of the dependence on the raw collocation matrix and the quadrature weights,
and part of the dependence on the overlap matrix, passes through the inverse
square root of the metric, whose response requires the solution of a
Sylvester equation and is treated below. The remaining dependence on the
overlap matrix is through the eigenvectors and eigenvalues in
Eq.~\eqref{eq:M} at fixed metric. The transfer matrix is invariant to any
orthogonal rotation of the retained canonically orthogonalized orbitals,
provided the same rotation is applied to $U_{pp'} s_{p'}^{-1/2}$ and
$U_{pp'} s_{p'}^{+1/2}$, so their response may be taken in any frame that
preserves $\sum_{pq} s_{p'}^{-1/2} U_{pp'} S_{pq} U_{qq'} s_{q'}^{-1/2} =
\delta_{p'q'}$. Taking the retained space to be geometry independent,
consistent with the treatment of the $\tau$ threshold in the underlying SCF
gradient, the symmetric choice is the matrix analogue of the derivative of an
inverse square root, with the inverse of the overlap matrix within the
retained space,
\begin{equation}
\nonumber
\pdiff{( U_{pp'} s_{p'}^{-1/2} )}{r_\xi}
=
-
\frac{1}{2}
\sum_{rs}
\left [
\sum_{q'}
U_{pq'}
s_{q'}^{-1}
U_{rq'}
\right ]
\end{equation}
\begin{equation}
\label{eq:frame_C}
\times
\pdiff{S_{rs}}{r_\xi}
U_{sp'}
s_{p'}^{-1/2}
\end{equation}
Since $U_{pp'} s_{p'}^{+1/2} = \sum_{q} S_{pq} U_{qp'} s_{p'}^{-1/2}$, the
dual factor follows by the product rule,
\begin{equation}
\nonumber
\pdiff{( U_{pp'} s_{p'}^{+1/2} )}{r_\xi}
=
\sum_{rs}
\left [
\delta_{pr}
-
\frac{1}{2}
\sum_{q'}
U_{pq'}
U_{rq'}
\right ]
\end{equation}
\begin{equation}
\times
\pdiff{S_{rs}}{r_\xi}
U_{sp'}
s_{p'}^{-1/2}
\end{equation}
where the bracket reduces to one half when no eigenvalues are discarded.
Differentiating Eq.~\eqref{eq:M} with these separates the response of the
transfer matrix into the part independent of the metric response and the
response of the inverse square root of the metric,
\begin{equation}
\label{eq:dM_split}
\sum_{pq}
\Pi_{pq}
\pdiff{M_{pq}}{r_\xi}
=
\sum_{pq}
\Sigma_{pq}^{(U)}
\pdiff{S_{pq}}{r_\xi}
+
\sum_{p'q'}
\tilde \Pi_{p'q'}
\pdiff{\Omega_{p'q'}^{-1/2}}{r_\xi}
\end{equation}
The second term contracts the pseudo-density of the inverse square root of
the metric, which follows directly from Eq.~\eqref{eq:M},
\begin{equation}
\nonumber
\tilde \Pi_{p'q'}
\equiv
\frac{1}{2}
\sum_{pq}
s_{p'}^{+1/2}
U_{pp'}
\Pi_{pq}
U_{qq'}
s_{q'}^{-1/2}
\end{equation}
\begin{equation}
+
\frac{1}{2}
\sum_{pq}
s_{p'}^{-1/2}
U_{pp'}
\Pi_{qp}
U_{qq'}
s_{q'}^{+1/2}
\end{equation}
and the first contains the part of the overlap pseudo-density that does not
require the Sylvester solution,
\begin{equation}
\nonumber
\Sigma_{pq}^{(U)}
\equiv
\frac{1}{2}
\sum_{p'q'}
\sum_{r}
s_{p'}^{-1/2}
U_{pp'}
\Omega_{p'q'}^{-1/2}
U_{rq'}
s_{q'}^{-1/2}
\Pi_{qr}
\end{equation}
\begin{equation}
\nonumber
+
\frac{1}{2}
\sum_{p'q'}
\sum_{r}
s_{p'}^{-1/2}
U_{qp'}
\Omega_{p'q'}^{-1/2}
U_{rq'}
s_{q'}^{-1/2}
\Pi_{pr}
\end{equation}
\begin{equation}
\nonumber
-
\frac{1}{2}
\sum_{p'q'}
\sum_{r'}
s_{p'}^{-1/2}
U_{pp'}
\Omega_{p'r'}^{-1/2}
\tilde \Pi_{r'q'}
U_{qq'}
s_{q'}^{-1/2}
\end{equation}
\begin{equation}
-
\frac{1}{2}
\sum_{p'q'}
\sum_{r'}
s_{p'}^{-1/2}
U_{pp'}
\tilde \Pi_{p'r'}
\Omega_{r'q'}^{-1/2}
U_{qq'}
s_{q'}^{-1/2}
\end{equation}
The remaining part of the overlap pseudo-density, together with
$\tilde \epsilon_{A}$ and $\tilde \Lambda_{p}^{A}$, follows from the metric
response.

The metric response is the last term of Eq.~\eqref{eq:dM_split},
\begin{equation}
\sum_{p'q'}
\tilde \Pi_{p'q'}
\pdiff{\Omega_{p'q'}^{-1/2}}{r_\xi}
\end{equation}
The derivative of the inverse square root of a matrix has no closed form in
terms of the derivative of the matrix, because the two do not commute.
Differentiating the identity $\sum_{r'} \Omega_{p'r'}^{-1/2}
\Omega_{r'q'}^{-1/2} = \Omega_{p'q'}^{-1}$ instead gives a Sylvester equation
(a linear matrix equation with the unknown multiplied from the left in one
term and from the right in the other) for it,
\begin{equation}
\nonumber
\sum_{r'}
\left [
\Omega_{p'r'}^{-1/2}
\pdiff{\Omega_{r'q'}^{-1/2}}{r_\xi}
+
\pdiff{\Omega_{p'r'}^{-1/2}}{r_\xi}
\Omega_{r'q'}^{-1/2}
\right ]
=
\pdiff{\Omega_{p'q'}^{-1}}{r_\xi}
\end{equation}
\begin{equation}
\label{eq:sylvester_forward}
=
-
\sum_{r's'}
\Omega_{p'r'}^{-1}
\pdiff{\Omega_{r's'}}{r_\xi}
\Omega_{s'q'}^{-1}
\end{equation}
Solving this equation separately for each atomic coordinate would require
the response of the metric for each, which is what the pseudo-density
formulation avoids. Instead, because only the contraction with
$\tilde \Pi_{p'q'}$ is needed, a single adjoint Sylvester equation with the
pseudo-density as its right-hand side is solved once,
\begin{equation}
\label{eq:sylvester_adjoint}
\sum_{r'}
\left [
\Omega_{p'r'}^{-1/2}
\Phi_{r'q'}
+
\Phi_{p'r'}
\Omega_{r'q'}^{-1/2}
\right ]
=
\tilde \Pi_{p'q'}
\end{equation}
Substituting this for $\tilde \Pi_{p'q'}$, using the symmetry of all of the
matrices involved to move the inverse square roots onto the derivative, and
recognizing the left-hand side of Eq.~\eqref{eq:sylvester_forward},
\begin{equation}
\nonumber
\sum_{p'q'}
\tilde \Pi_{p'q'}
\pdiff{\Omega_{p'q'}^{-1/2}}{r_\xi}
=
\sum_{p'q'}
\Phi_{p'q'}
\pdiff{\Omega_{p'q'}^{-1}}{r_\xi}
\end{equation}
\begin{equation}
\label{eq:metric_response}
=
-
\sum_{p'q'}
\sum_{r's'}
\Phi_{p'q'}
\Omega_{p'r'}^{-1}
\pdiff{\Omega_{r's'}}{r_\xi}
\Omega_{s'q'}^{-1}
\end{equation}
The response of the inverse square root of the metric has been traded for
the response of its inverse, which is available in closed form, at the cost
of one Sylvester solve that is independent of the atomic coordinate. With
the eigendecomposition of the metric, which is already available from the
construction of $\Omega_{p'q'}^{-1/2}$ and whose eigenbasis is labeled by
double-primed indices,
\begin{equation}
\Omega_{p'q'}
\equiv
\sum_{p''}
\Theta_{p'p''}
\theta_{p''}
\Theta_{q'p''}
\end{equation}
the adjoint equation is diagonal in the eigenbasis and its solution is,
\begin{equation}
\label{eq:Phi}
\Phi_{p'q'}
=
\sum_{p''q''}
\Theta_{p'p''}
\frac{
\sum_{r's'}
\Theta_{r'p''}
\tilde \Pi_{r's'}
\Theta_{s'q''}
}{
\theta_{p''}^{-1/2}
+
\theta_{q''}^{-1/2}
}
\Theta_{q'q''}
\end{equation}
which is symmetric and involves no small denominators for a positive
definite metric.

By Eq.~\eqref{eq:Omega}, the metric depends on the atomic positions through
the quadrature weights and the raw collocation matrix and, through the
orthogonalization factors in the symmetric frame of Eq.~\eqref{eq:frame_C},
through the overlap matrix, so Eq.~\eqref{eq:metric_response} splits into
three contractions,
\begin{equation}
\nonumber
-
\sum_{p'q'}
\sum_{r's'}
\Phi_{p'q'}
\Omega_{p'r'}^{-1}
\pdiff{\Omega_{r's'}}{r_\xi}
\Omega_{s'q'}^{-1}
=
\sum_{A}
\tilde \epsilon_{A}
\pdiff{w_{A}}{r_\xi}
\end{equation}
\begin{equation}
\label{eq:metric_response_split}
+
\sum_{A}
\sum_{p}
\tilde \Lambda_{p}^{A}
\pdiff{X_{p}^{A}}{r_\xi}
+
\sum_{pq}
\Sigma_{pq}^{(\Omega)}
\pdiff{S_{pq}}{r_\xi}
\end{equation}
in which $\Phi_{p'q'}$ appears transformed in three different ways. For the
pseudo-energy density it is sandwiched between inverse metrics and contracted
on both sides with the collocation matrix of the canonically orthogonalized
orbitals,
\begin{equation}
\nonumber
\tilde \epsilon_{A}
\equiv
-
\sum_{p'q'}
\left [
\sum_{p}
s_{p'}^{-1/2}
U_{pp'}
X_{p}^{A}
\right ]
\end{equation}
\begin{equation}
\label{eq:eps_tilde}
\times
\left [
\sum_{r's'}
\Omega_{p'r'}^{-1}
\Phi_{r's'}
\Omega_{s'q'}^{-1}
\right ]
\left [
\sum_{q}
s_{q'}^{-1/2}
U_{qq'}
X_{q}^{A}
\right ]
\end{equation}
for the collocation pseudo-density the same kernel is contracted with that
collocation matrix on one side and carried back to the atomic orbital basis
on the other, with a factor of two from the two collocation matrices in the
metric,
\begin{equation}
\nonumber
\tilde \Lambda_{p}^{A}
\equiv
-
2
w_{A}
\sum_{p'q'}
s_{p'}^{-1/2}
U_{pp'}
\left [
\sum_{r's'}
\Omega_{p'r'}^{-1}
\Phi_{r's'}
\Omega_{s'q'}^{-1}
\right ]
\end{equation}
\begin{equation}
\label{eq:Lambda_tilde}
\times
\left [
\sum_{q}
s_{q'}^{-1/2}
U_{qq'}
X_{q}^{A}
\right ]
\end{equation}
and for the metric part of the overlap pseudo-density one inverse metric is
cancelled by the metric in the frame derivative, leaving a symmetrized
product with a single inverse metric carried back to the atomic orbital basis
on both sides,
\begin{equation}
\nonumber
\Sigma_{pq}^{(\Omega)}
\equiv
\frac{1}{2}
\sum_{p'q'}
\sum_{r'}
s_{p'}^{-1/2}
U_{pp'}
\Omega_{p'r'}^{-1}
\Phi_{r'q'}
U_{qq'}
s_{q'}^{-1/2}
\end{equation}
\begin{equation}
+
\frac{1}{2}
\sum_{p'q'}
\sum_{r'}
s_{p'}^{-1/2}
U_{pp'}
\Phi_{p'r'}
\Omega_{r'q'}^{-1}
U_{qq'}
s_{q'}^{-1/2}
\end{equation}
Restoring the fixed-metric part of Eq.~\eqref{eq:dM_split} gives the
complete response of the transfer matrix,
\begin{equation}
\nonumber
\sum_{pq}
\Pi_{pq}
\pdiff{M_{pq}}{r_\xi}
=
\sum_{A}
\tilde \epsilon_{A}
\pdiff{w_{A}}{r_\xi}
+
\sum_{A}
\sum_{p}
\tilde \Lambda_{p}^{A}
\pdiff{X_{p}^{A}}{r_\xi}
\end{equation}
\begin{equation}
\label{eq:dM_final}
+
\sum_{pq}
\left [
\Sigma_{pq}^{(U)}
+
\Sigma_{pq}^{(\Omega)}
\right ]
\pdiff{S_{pq}}{r_\xi}
\end{equation}
which identifies the overlap pseudo-density of Eq.~\eqref{eq:dM_decomp} as
$\Sigma_{pq} = \Sigma_{pq}^{(U)} + \Sigma_{pq}^{(\Omega)}$.

Inserting Eq.~\eqref{eq:dM_final} into Eq.~\eqref{eq:cgridk_grad} and
collecting the contractions against each integral derivative gives the
overlap-fitted Grid-K exchange gradient,
\begin{equation}
\nonumber
\bar E_{K}^{\mathrm{grid}, \xi}
[ D_{pq} ]
=
\sum_{A}
\left [
\bar \epsilon_{A}
+
\tilde \epsilon_{A}
\right ]
\pdiff{w_{A}}{r_\xi}
\end{equation}
\begin{equation}
\nonumber
+
\sum_{A}
\sum_{p}
\left [
\Lambda_{p}^{A}
+
\tilde \Lambda_{p}^{A}
\right ]
\pdiff{X_{p}^{A}}{r_\xi}
\end{equation}
\begin{equation}
\nonumber
+
\sum_{A}
\sum_{pq}
\bar \Gamma_{pq}^{A}
\pdiff{V_{pq}^{A}}{r_\xi}
\end{equation}
\begin{equation}
\label{eq:cgridk_grad_final}
+
\sum_{pq}
\left [
\Sigma_{pq}^{(U)}
+
\Sigma_{pq}^{(\Omega)}
\right ]
\pdiff{S_{pq}}{r_\xi}
\end{equation}
The structure is that of the unfitted gradient of
Eq.~\eqref{eq:gridk_grad} with augmented pseudo-densities and one additional
contraction, against overlap derivative integrals that the SCF gradient
already requires. The potential integral derivative contraction, which
dominates the cost of the unfitted gradient, is unchanged in form and in
count. For $N_{\mathrm{AO}}$ atomic orbitals, $N_{\mathrm{MO}}$ retained
canonically orthogonalized orbitals, and $N_{\mathrm{grid}}$ grid points, the
additional work is the grid contraction forming $\Pi_{pq}$ in
Eq.~\eqref{eq:Pi}, of the same $\mathcal{O}(N_{\mathrm{AO}}^{2}
N_{\mathrm{grid}})$ type as the formation of the metric, the
$\mathcal{O}(N_{\mathrm{AO}} N_{\mathrm{MO}} N_{\mathrm{grid}})$ grid-side
products in Eqs.~\eqref{eq:eps_tilde} and \eqref{eq:Lambda_tilde}, and
$\mathcal{O}(N_{\mathrm{AO}}^{2} N_{\mathrm{MO}})$ dense linear algebra,
including one Sylvester solve in the eigenbasis of the metric that is already
available from the overlap-fitted exchange matrix. No new integral derivative
types are introduced.

When using grid-based exchange algorithms, it is always tempting to converge
the SCF equations using a coarse exchange integration grid and introduce post 
hoc corrections to improve the accuracy. This is commonly done by re-evaluating
the exchange energy using a denser grid\cite{Neese:2009:98,Izsak:2011:144105,Bochevarov:2013:2110}
or by introducing analytic corrections that depend on the converged density.\cite{Zhang:2025:1492}
While this may provide improved accuracy in the energy, it makes evaluation of the analytic gradient 
challenging -- although it remains perfectly well-defined. Strictly speaking,
the analytic gradient of the post hoc corrected energy requires the solution
of coupled-perturbed SCF equations as well as evaluation of ``skeleton'' contributions\cite{Yamaguchi:1994:NewDimension}
from the correction. If the true analytic gradient was evaluated, one arrives at a ``no free lunch'' scenario, where
the increased cost of an analytic gradient likely offsets any gains in efficiency realized
from the correction. In practice, approximations to the true gradient may be 
suitable for many applications (as in the case of re-evaluating the energy
using a denser grid). However, if results in larger grids are desired, our recommendation is to switch grids during the SCF
procedure and converge the equations using the larger grid, thus avoiding any
increased cost during the evaluation of the analytic gradient and thus avoiding
the need for additional approximations. A related issue
arises in one-sided overlap-fitted COSX,\cite{Izsak:2011:144105} where the fit
breaks the permutational symmetry of the approximate ERI tensor. A variational
treatment requires that the tensor itself be symmetrized, which amounts to a
second density contraction with the roles of the fitted and unfitted
collocation matrices exchanged. Symmetrizing the exchange matrix a posteriori
instead yields the correct exchange energy but an exchange matrix that is not
its derivative with respect to the density matrix, so the converged density is
not stationary with respect to the energy and the SCF energy is no longer
variational. Such issues can be avoided entirely by using a symmetric
factorization such as the one presented here or the semiJK method of Holzer.\cite{Holzer:2020:184115}

\subsection{Compensated Grid-Based Exchange}

The dSCF and overlap-fitting procedures described above are orthogonal
considerations, and can be separately applied with Grid-K to yield different
methods. We refer to the union of the two methods, which we believe provides
compelling accuracy vs cost properties, as ``compensated grid-based exchange''
or compensated Grid-K. 

\section{Implementation}

All stages of the compensated Grid-K method above are implementable in v0.3 of
the cuEST CUDA-X library for quantum chemistry. Utilities for Grid-K,
symmetrically overlap-fitted extensions thereof, and gradients thereof for
arbitrary symmetric density matrix inputs are provided as primary cuEST API
functions. An example of the dSCF procedure including use of the ANO-DK3 basis\cite{Tsuchiya:2001:4463}
projected into the current basis to form the SAD density matrix and use of the
dSCF procedure within a compensated Grid-K SCF procedure are provided as code
samples in the simple \texttt{cuest\_scf} case study. Analytical SAD exchange
matrices and gradients are computed with the new JIT-based integral direct APIs
in cuEST v0.3, and this integral direct API is also used for the exact reference
energy values below. Density fitting (e.g., for the Coulomb matrix) is performed
where indicated below using cuEST API functions in the def2-universal-JKFIT\cite{Weigend:2008:167}
basis unless otherwise indicated. All computations are performed in FP64-quality
numerics unless otherwise indicated. 

\subsection{SAD K Primitives}

To maximize the accelerations possible with compensated Grid-K, it is useful to
have a small library of efficient methods for the manipulation of exchange
operators with SAD density matrices. The following primitives are particularly
useful:

SAD reference system hybrid exchange energy,
\begin{equation}
E_{K}^{\varnothing} [ D_{pq}^{\varnothing} ]
\equiv
\frac{1}{2}
\sum_{pq}
\sum_{rs}
D_{pq}^{\varnothing}
(pr|\hat o|qs)
D_{rs}^{\varnothing}
\end{equation}
SAD reference system hybrid exchange energy atomic gradient,
\begin{equation}
E_{K}^{\varnothing, \xi} [ D_{pq}^{\varnothing} ]
\equiv
\frac{1}{2}
\sum_{pq}
\sum_{rs}
D_{pq}^{\varnothing}
\pdiff{
(pr|\hat o|qs)
}{
r_\xi
}
D_{rs}^{\varnothing}
\end{equation}
SAD reference system external potential,
\begin{equation}
K_{pq}^{\varnothing} [D_{rs}^{\varnothing}]
\equiv
\sum_{rs}
(pr|\hat o|qs)
D_{rs}^{\varnothing}
\end{equation}
SAD reference system external potential bilinear atomic gradient,
\begin{equation}
E_{K}^{\varnothing \Delta, \xi}
[D_{pq}^{\varnothing}, D_{pq}^{\Delta}]
\equiv
\sum_{pq}
\sum_{rs}
D_{pq}^{\varnothing}
\pdiff{
(pr|\hat o|qs)
}{
r_\xi
}
D_{rs}^{\Delta}
\end{equation}
It is usually best if this last primitive makes no assumptions about the
structure of $D_{rs}^{\Delta}$.

All of these terms can be obtained with reasonable efficiency from a
high-quality general-purpose integral-direct exchange code, which we exploit
herein. However, even higher performance is possible with specialization given
the advantageous matrix sparsity, physical locality, and sphericality properties
of $D_{pq}^{\varnothing}$. 

\subsection{Compensated Grid-K Primitives}

The implementation of Grid-K follows the equations presented earlier quite closely.
The collocation matrix is formed using CUDA kernels (Eq.~\eqref{eq:collocation}),
the collocation matrix is multiplied by the density using GEMM routines from 
cuBLAS\cite{cuBLAS} (Eq.~\eqref{eq:gridk_Y}), and the rate-limiting step is the partial contraction of
the $Y$ intermediate against the potential integrals (Eq.~\eqref{eq:gridk_Z}).
The formation of $Z$ is a series of kernels and can be implemented using any standard
Boys or Rys-based techniques for evaluating the Coulomb operator. Finally, 
another GEMM assembles the exchange matrix (Eq.~\eqref{eq:gridk_K}). This procedure is
blocked by grid points to control memory usage and sparsity of the AO basis is exploited
to reduce the overall scaling. For symmetric input densities, the exchange matrix is 
symmetrized; for nonsymmetric input densities, the procedure is performed twice:
once for the density and once for the transposed density. The implementation of the
gradient follows similarly, albeit with a few extra kernels required for potential
integral derivatives and derivative collocation matrices (Eq.~\eqref{eq:gridk_grad}).

Overlap fitting of Grid-K does not require any new integral primitives and only a few new
linear algebra operations. The overlap matrix must be formed by quadrature
($S^{\rm grid}$). This requires only the usual orbital collocation primitives
and a few GEMM operations.  To form the transfer matrix, we diagonalize the
overlap matrix and the grid overlap in the orthogonal basis. We use
cuSOLVER\cite{cuSOLVER} for these operations. Applying the overlap fitting in
Grid-K is simply a matter of transforming the density prior to the usual Grid-K
algorithm and transforming the resultant exchange matrix.  Again, the gradient
is somewhat more involved, but no new primitives are needed; it is simply a
matter of additional linear algebra.

\subsection{Becke Grid Selection}

All grids used in this work are of the ubiquitous Becke molecular quadrature
type. For each atom-centered grid, a pruned grid composed of a radial quadrature
rule with varying Lebedev solid angle
quadratures\cite{Lebedev:1976:10,Lebedev:1999:477} at each radial node is used.
The direct product of the radial quadrature weight (including the $r^2$ Jacobian
factor) and the angular quadrature weight gives the total raw atomic quadrature
weight $v_{A} \equiv v_{A}^{R} v_{A}^{\Omega}$. Without further adjustment,
these $N_{\mathrm{atom}}$ atom-centered quadrature grids would integrate
$\mathbb{R}^3$ $N_{\mathrm{atom}}$ times, and would have poor support for
transients in integrands far from each atomic center. To avoid this, the
integrands are partitioned to atom-centered basins according to a smooth Voronoi
decomposition function $u_{A}$,\cite{Becke:1988:2547} giving the total finished quadrature weight of
$w_{A} \equiv u_{A} v_{A}$. In particular, we use the $C^{3}$-smooth Stratmann
switching function extension\cite{Stratmann:1996:213} of the original Becke smooth Voronoi decomposition,
without atomic size adjustments. In this work, all atomic grids are fixed to the
lab frame, meaning that the atomic quadrature positions advect rigidly with the
atomic centers in this treatment. In more elaborate schemes where the atomic
grids are oriented to the molecular or local inertial
frames,\cite{Gill:1993:506,Johnson:1994:377} the grid points
would also rotate with the chosen frame.  Finally, the Becke--Stratmann weights
depend on the atomic positions.  All of these effects induce geometric cross
terms in analytical gradients, which we faithfully support in all gradient
implementations shown herein. 

Throughout this work, we use the Standard Grids
(SG)\cite{Gill:1993:506,Chien:2006:730,Dasgupta:2017:869} for exchange--correlation
(Vxc), nonlocal correlation (VV10), and Grid-K hybrid
exchange (K). Specifically, we use SG-2 for Vxc, SG-0 for VV10, and SG-0 or SG-1
as noted for Grid-K.

\section{Results}

\subsection{Quick Tour of Compensated Grid-K}

\textbf{\textit{Note:}} There is an immense hypercube of considerations around
compensated Grid-K: system size, DFT method, basis set quality and/or
augmentation, grid size, choice of dSCF and/or symmetric overlap fitting
compensations, and accuracy of absolute or relative properties. This section is
designed to quickly explore some aspects of this hypercube, while later results
sections will provide more rigorous benchmarks of certain dimensions. 

\subsubsection{Starting point - Gramicidin Dimer \texorpdfstring{$\omega$}{ω}B97M-V/def2-TZVP with
SG-0 for Compensated Grid-K}

\begin{figure*}[p]
\centering
\includegraphics[width=\textwidth]{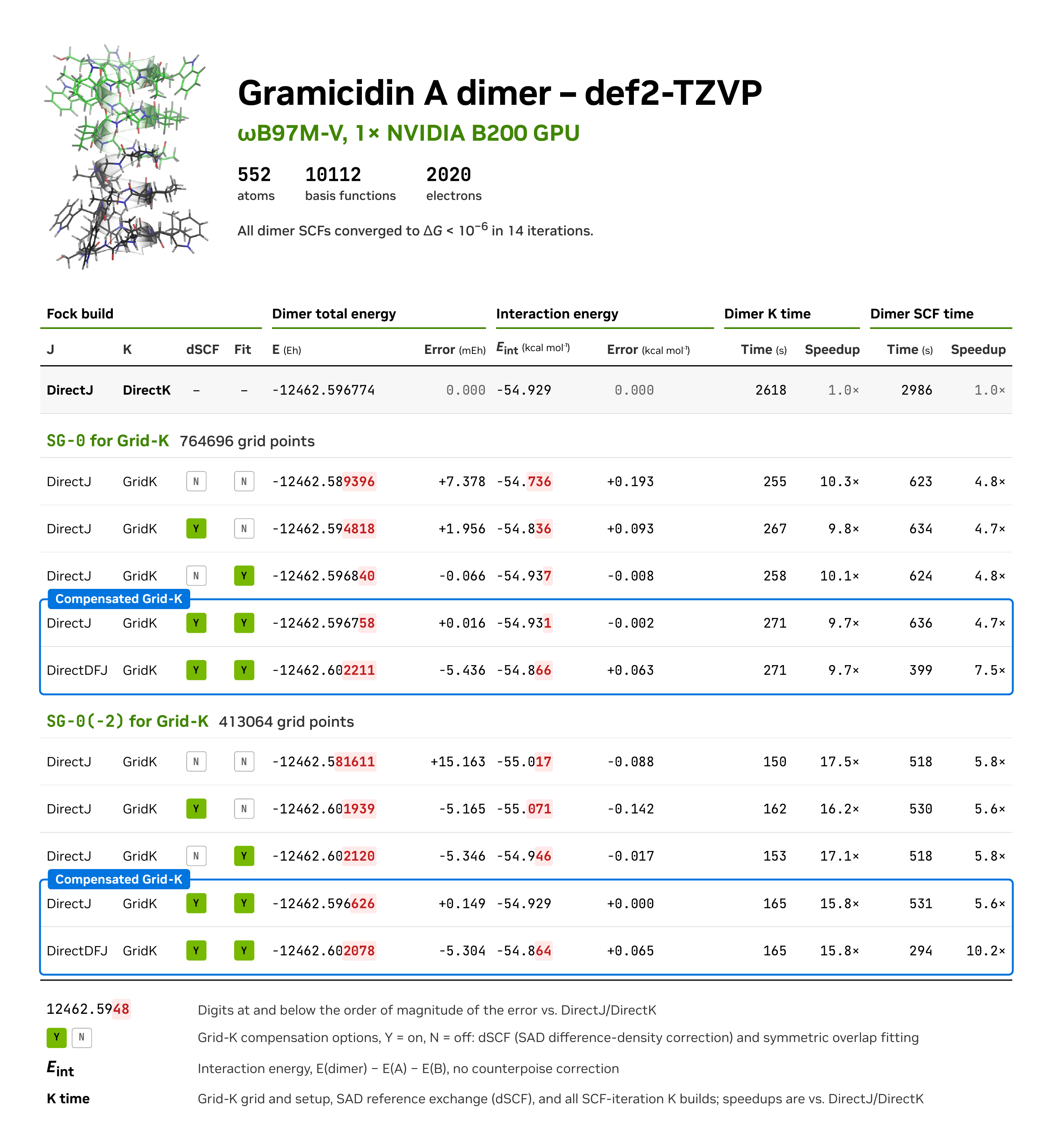}
\caption{Gramicidin A dimer $\omega$B97M-V/def2-TZVP compensated Grid-K case
study, with Grid-K on the SG-0 (top) and SG-0($-2$) (bottom) grids.}
\label{fig:tour_gA_tzvp}
\end{figure*}

Consider the gramicidin A head-to-head channel dimer (PDB 1JNO, Figure
\ref{fig:tour_gA_tzvp}),\cite{Townsley:2001:11676} a neutral singlet homo-dimer
with a total of 552 atoms. This system is chosen as a spatially compact geometry
with an interesting intermolecular interaction motif in the $\sim$500 atom size
regime.

Consider the def2-TZVP basis, which has 10112 atomic orbitals for this system
and includes up to F functions. The $\omega$B97M-V/def2-TZVP energy, e.g., as
evaluated with DirectJ/K integrals is $-12462.596774$ H. Note that m$E_{\rm h}$ precision
is $\sim 10^{-7}$ relative precision, uH precision is $\sim 10^{-10}$ relative
precision.  Grid-K in the SG-0 grid yields an absolute error of 7.4 m$E_{\rm h}$. Grid-K
with a dSCF correction fares better, with absolute error of 2.0 m$E_{\rm h}$. Grid-K with
symmetric overlap fitting is the first to fall below m$E_{\rm h}$ absolute error, and
yields an error of 66 uH. Compensated Grid-K (with both dSCF and symmetric
overlap fitting) yields an error of 16 uH. Note that all of these energies are
differentiable.

The same hierarchy holds for the interaction energy between the two monomers,
which is $-54.93$ kcal mol$^{-1}$ with DirectJ/K integrals (monomers frozen at
the dimer geometry, no counterpoise correction). Error cancellation between the
dimer and the monomers already helps uncompensated Grid-K, which is in error by
0.19 kcal mol$^{-1}$; the dSCF correction reduces this to 0.093 kcal mol$^{-1}$,
symmetric overlap fitting to 0.008 kcal mol$^{-1}$, and compensated Grid-K to
0.002 kcal mol$^{-1}$. All variants converge in the same number of SCF
iterations (14): the dimer computation with DirectJ/K takes 2986 s on a single
NVIDIA B200 GPU, whereas Grid-K takes 623 s (a 4.8$\times$ end-to-end
SCF speedup) and compensated Grid-K takes 636 s (4.7$\times$ end-to-end SCF
speedup).  Replacing the direct Coulomb build with DF-J reduces the time further
to 399 s (7.5$\times$ end-to-end SCF speedup), at the cost of errors of 5.4 m$E_{\rm h}$
in the total energy and 0.063 kcal mol$^{-1}$ in the interaction energy that are
attributable to the density fitting of the Coulomb term rather than to Grid-K.
Speedups for the K term in isolation (including Grid-K setup) are even higher,
with 9.7$\times$ realized for our preferred compensated Grid-K approach. 

\subsubsection{Smaller grids - the SG-0(\texorpdfstring{$-\Delta$}{−Δ}) series}

The observant reader may have noticed that compensated Grid-K with SG-0 appears
to be a bit overgunned for this problem, with absolute and relative accuracy
properties that exceed the well-trusted DF-J approximation. Is there a
possibility to use smaller grids with compensated Grid-K and realize an even
faster method with similar error properties as DF-J? The answer seems to be
preliminarily yes, for two reasons: (1) there is a long history of use of
smaller grids than SG-0 in fitted flavors of Grid-K throughout the literature
and (2) we here construct an extremely brief example that shows that grids with
roughly one half the number of points of SG-0 may be safely used for these
problems. However, we note that a full study of smaller recommended grids than
SG-0 lies outside the scope of this paper. 

While a proper study of smaller grids would optimize such grids for the problem
at hand, we will fall back to a simpler heuristic. Following a procedure we once
used in ab initio scattering computations,\cite{Parrish:2019:1523} we define the
SG-0($-\Delta$) family of trimmed SG-0 grids. Each $\Delta$ level is defined by
trimming the Lebedev grids at all radial nodes in SG-0 $\Delta$ times along the
sequence $[170, 146, 110, 86, 74, 50, 38, 26, 18, 6]$ (note that the nonstandard
18-point Lebedev grid is used in SG-0\cite{Chien:2006:730} rather than the usual 14-point Lebedev
grid). When the Lebedev grid at a given radial node makes the terminal trimming
step from the 6-point Lebedev grid to zero, we delete the radial node and step
down to the MultiExp\cite{Gill:2003:732,Parrish:2023:2305.01621} radial grid
with one fewer node (so all of the remaining radial
nodes adjust in position). Note that radial node deletions do not change the
ordering of the remaining radial nodes, and the atomic size parameter does not
change during this procedure.

For the gramicidin A dimer def2-TZVP study of the previous subsection,
SG-0($-2$) (bottom block of Figure \ref{fig:tour_gA_tzvp}) yields pragmatically
similar accuracy as SG-0 while bringing the K term speedup to $15.8\times$ and
end-to-end SCF speedup to $10.2\times$ when used with DirectDF-J.  Marginally
higher total energy errors are observed with SG-0($-2$), but these cancel well
in interaction energies, where errors are of the same order as in SG-0. 

\subsubsection{Larger basis sets - def2-QZVP}

Consider the gramicidin dimer from above, but now with the def2-QZVP
basis, which has 23904 atomic orbitals for this system and includes up to G
functions.

\begin{figure*}[p]
\centering
\includegraphics[width=\textwidth]{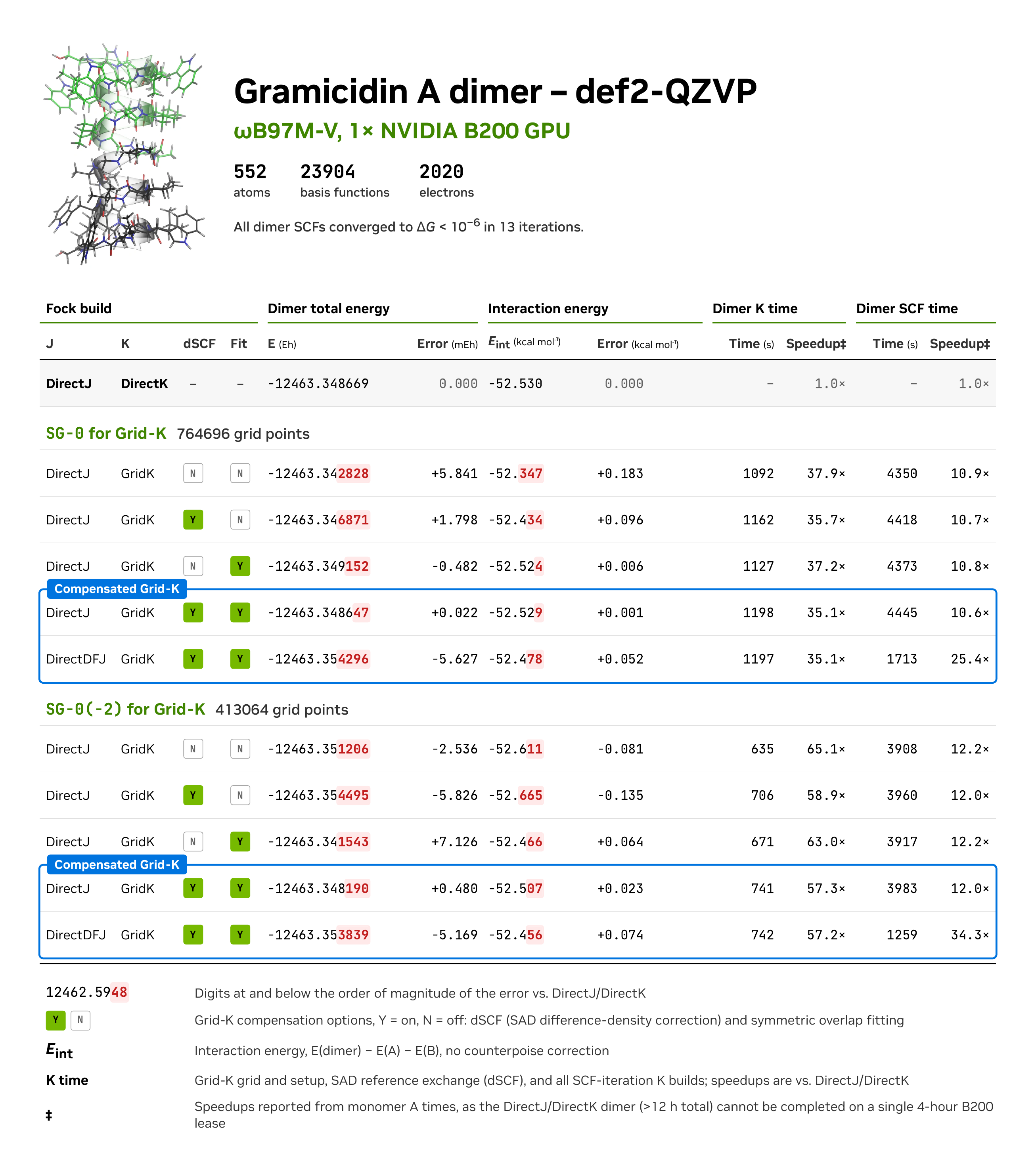}
\caption{Gramicidin A dimer $\omega$B97M-V/def2-QZVP compensated Grid-K case
study, with Grid-K on the SG-0 (top) and SG-0($-2$) (bottom) grids. Speedups (\ddag) are
measured on monomer A.}
\label{fig:tour_gA_qzvp}
\end{figure*}

Overall, compensated Grid-K errors in SG-0 (Figure \ref{fig:tour_gA_qzvp}) are
not qualitatively distinguishable
from those in def2-TZVP. More detailed studies on smaller systems pursued in
later sections indicate that the error behavior of def2-QZVP may be even more
favorable than def2-TZVP. Speedups of compensated Grid-K are dramatic (measured
on monomer A, as the DirectJ/DirectK dimer, $>12$ h, cannot be completed on a
single 4-hour B200 lease): In SG-0,
K term speedups are $35.1\times$, yielding $10.6\times$ end-to-end SCF speedup
in DirectJ and $25.4\times$ end-to-end SCF speedup in DirectDF-J.

SG-0($-2$) absolute and relative errors remain excellent, with a compensated
Grid-K interaction energy error of $0.023$ kcal mol$^{-1}$.  In SG-0($-2$), K term
speedups are $57.2\times$, yielding $12.0\times$ end-to-end SCF speedup in
DirectJ and $34.3\times$ end-to-end SCF speedup in DirectDF-J.

\subsubsection{Larger molecules - beyond 1000 atoms in def2-TZVP}

Consider the GNNQQNY steric zipper from the yeast prion Sup35 (PDB
1YJP),\cite{Nelson:2005:773} a neutral singlet homo-dimer of two five-strand
$\beta$-sheets with a total of 1070 atoms. This system is chosen as a spatially
compact geometry with an interesting intermolecular interaction motif in the
$\sim$1000 atom size regime. The def2-TZVP basis has 21170 atomic orbitals for
this system.

Note that we are temporarily limited in cuEST in that we do not yet have a
Krylov solve method for DirectDF-J, and therefore cannot yet afford the metric
pseudoinversion (``Poisson solve'') step for systems of this size. Therefore no
DirectDF-J results are shown, which will cause DirectJ to limit potential
end-to-end SCF speedups. 

\begin{figure*}[p]
\centering
\includegraphics[width=\textwidth]{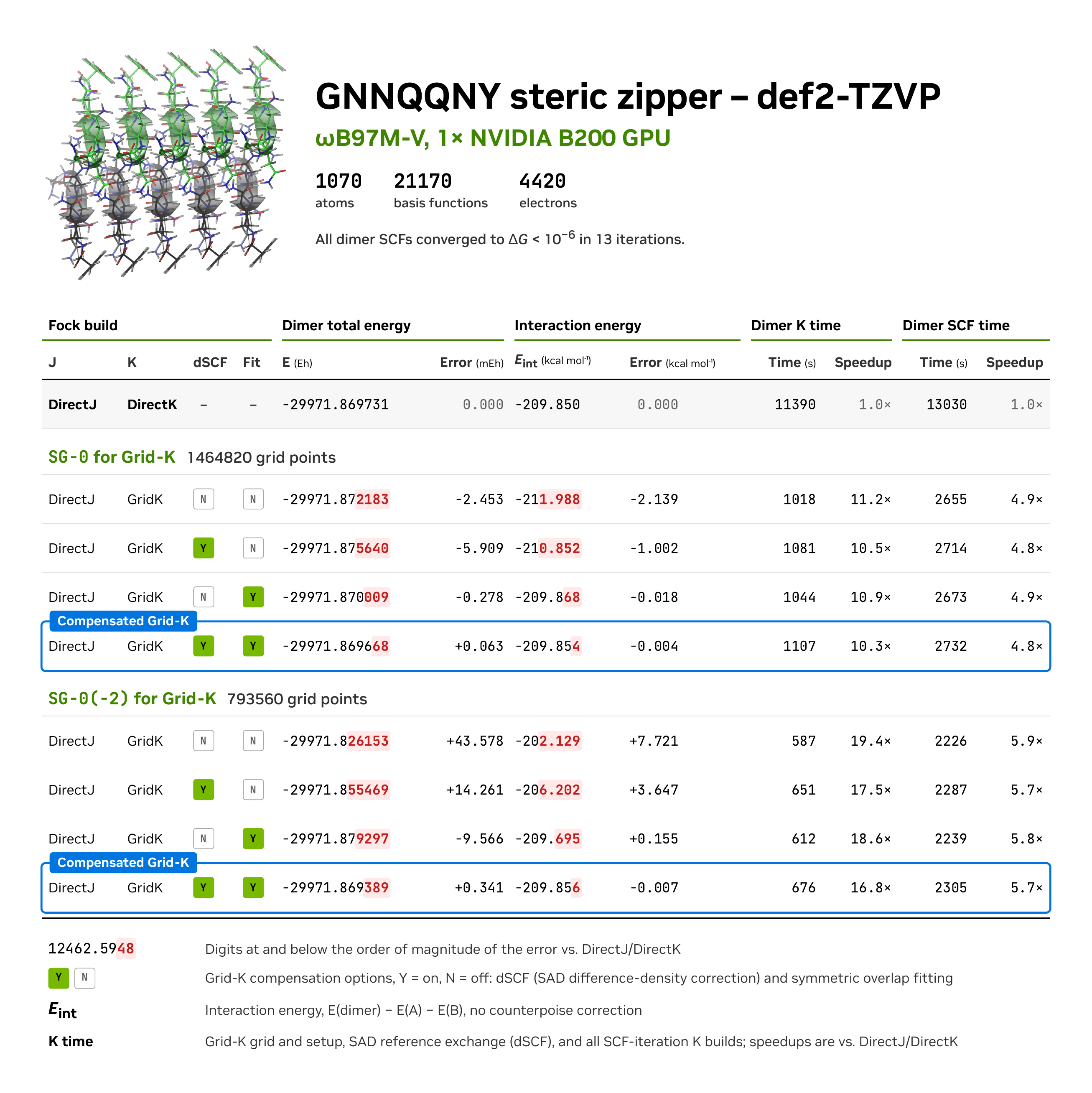}
\caption{GNNQQNY steric zipper $\omega$B97M-V/def2-TZVP compensated Grid-K case
study, with Grid-K on the SG-0 (top) and SG-0($-2$) (bottom) grids.}
\label{fig:tour_gn_tzvp}
\end{figure*}

Interaction energy accuracies are excellent in compensated Grid-K (Figure
\ref{fig:tour_gn_tzvp}), with 0.004 and 0.007 kcal
mol$^{-1}$ errors obtained in SG-0 and SG-0($-2$), respectively. Note that for
this large interaction energy motif, compensation dramatically improves the
interaction energy from the 2.1 and 7.7 kcal mol$^{-1}$ errors obtained with
uncompensated Grid-K in SG-0 and SG-0($-2$), respectively. Speedups are slightly
higher than the smaller gramicidin A dimer case - K term speedups are
$10.3\times$ and $16.8\times$ in SG-0 and SG-0($-2$), respectively. End-to-end SCF
speedups are $4.8\times$ and $5.7\times$, in SG-0 and SG-0($-2$) with DirectJ,
respectively, with overall speedup limited by a lack of tractability of
DirectDF-J.

One important callout is the routine tractability of compensated Grid-K for
systems of this scale:
\begin{itemize}
\item Gramicidin A dimer $\omega$B97M-V/def2-TZVP (10112 nao) requires 294 s for
end-to-end SCF (DirectDF-J, compensated Grid-K with SG-0($-2$)) with $<$0.1 kcal
mol$^{-1}$ error in the interaction energy stemming mostly from DF-J. 
\item Gramicidin A dimer $\omega$B97M-V/def2-QZVP (23904 nao) requires 1259 s for
end-to-end SCF (DirectDF-J, compensated Grid-K with SG-0($-2$)) with $<$0.1 kcal
mol$^{-1}$ error in the interaction energy stemming mostly from DF-J. 
\item GNNQQNY steric zipper $\omega$B97M-V/def2-TZVP (21170 nao) requires 2305 s for
end-to-end SCF (DirectJ, compensated Grid-K with SG-0($-2$)) with $<$0.01 kcal
mol$^{-1}$ error in the interaction energy from compensated Grid-K. (This size
regime is waiting on Krylov solver tractability increase for DirectDF-J, at
which point further speedup is possible). 
\end{itemize}
All of these results are obtained on $1\times$ B200 GPU. All of these
observables are differentiable. 

\subsection{Absolute Property Accuracy}

\begin{figure*}[p]
\centering
\includegraphics[width=0.85\textwidth]{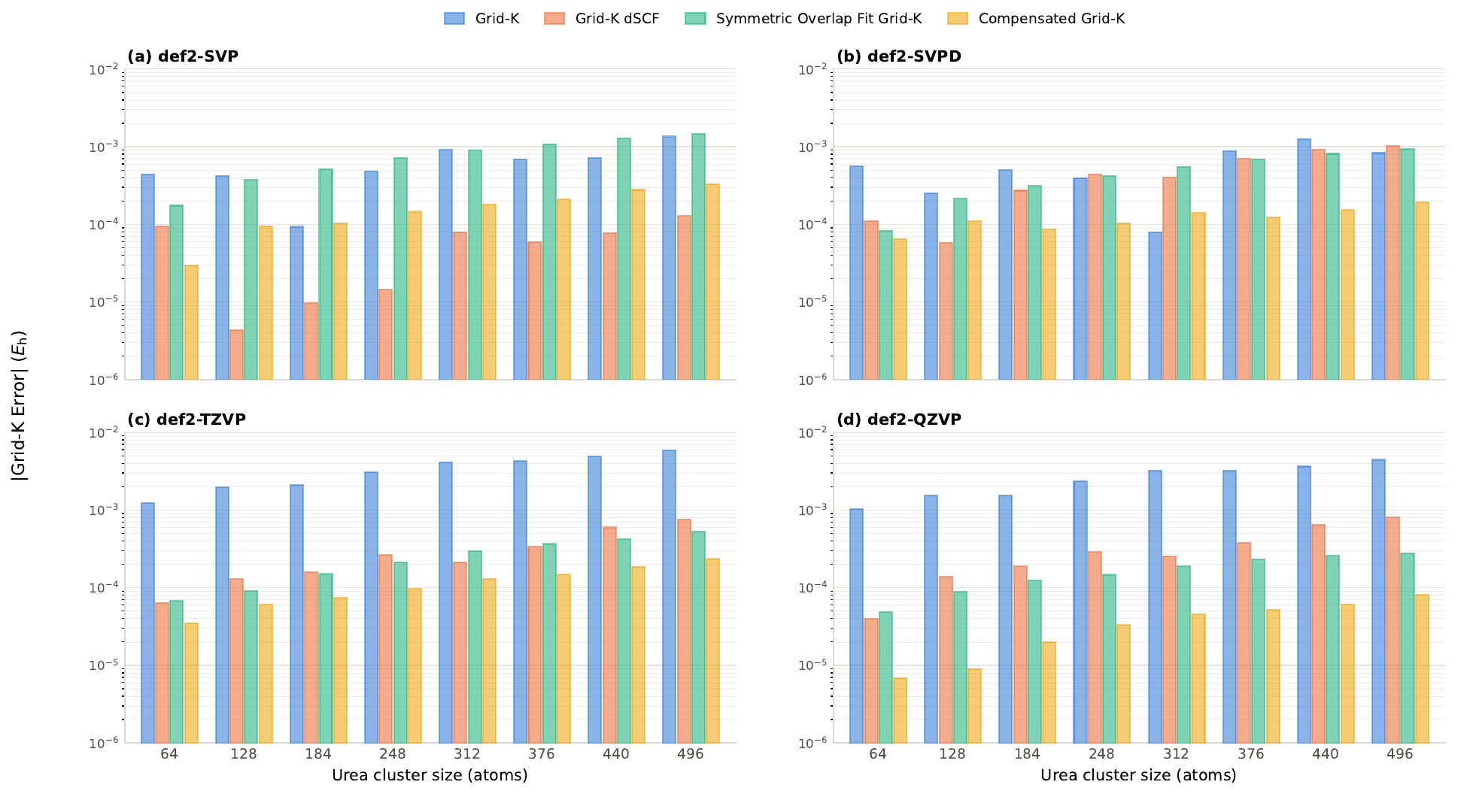}
\caption{Magnitude of the signed error in the $\omega$B97M-V total energy,
$|E^{\mathrm{Grid\text{-}K}} - E^{\mathrm{analytic}}|$, on a logarithmic scale,
for clusters of 8 to 62 urea molecules (64 to 496 atoms) cut from the urea
crystal structure, computed with canonical Grid-K, Grid-K with the SAD
correction (dSCF), the symmetric orthogonal overlap fit, and compensated
Grid-K on the SG-0 exchange grid in the (a) def2-SVP, (b) def2-SVPD, (c)
def2-TZVP, and (d) def2-QZVP basis sets. The errors are measured against
total energies computed with analytic two-electron integrals at the same
functional and basis set.}
\label{fig:abs_error}
\end{figure*}

Total energies offer no cancellation of the grid error between similar
structures and therefore expose the error of each Grid-K variant directly,
along with its growth with system size. We computed $\omega$B97M-V total
energies for a sequence of clusters cut from the experimental urea crystal
structure (COD entry 1008775),\cite{Guth:1980:237} containing 8 to 62 urea
molecules (64 to 496 atoms), in the def2-SVP, def2-SVPD, def2-TZVP, and def2-QZVP basis sets with
each Grid-K variant on the SG-0 exchange grid, and compared them with total
energies computed with analytic two-electron integrals. Figure \ref{fig:abs_error} shows the
magnitude of the signed error in the total energy,
$E^{\mathrm{Grid\text{-}K}} - E^{\mathrm{analytic}}$, on a logarithmic scale
for each cluster and basis set. The error of canonical Grid-K grows linearly
with the number of atoms in every basis set and is largest in the triple- and
quadruple-zeta sets, reaching 5.9 m$E_\mathrm{h}$ in
def2-TZVP and 4.4 m$E_\mathrm{h}$ in def2-QZVP for the
496-atom cluster. The two ingredients of the compensation scheme act
differently across the basis sets: the symmetric orthogonal overlap fit
reduces the error by an order of magnitude in def2-TZVP and def2-QZVP but
hardly changes it in the two double-zeta sets, whereas the SAD correction
reduces it by roughly an order of magnitude in every basis set. The
basis-set dependence of the overlap fit is the trend anticipated in the
Introduction: the fit can only correct components of the orbital products
that lie within the span of the orbital basis, so it is most effective in the
polarized triple- and quadruple-zeta sets, consistent with Holzer's
observation for the de-aliasing procedure of semiJK.\cite{Holzer:2020:184115}
In def2-TZVP and def2-QZVP, compensated Grid-K yields the smallest error for every
cluster. In the double-zeta basis sets the variants without the overlap fit
are occasionally smaller in magnitude, but their residuals change sign from
cluster to cluster, whereas the compensated Grid-K error is systematic:
positive in def2-SVP, def2-TZVP, and def2-QZVP, negative in def2-SVPD, and
growing linearly with cluster size. For the largest cluster the compensated
Grid-K error decreases with basis-set quality, from 0.33 m$E_\mathrm{h}$
in def2-SVP and 0.19 m$E_\mathrm{h}$ in def2-SVPD to
0.23 m$E_\mathrm{h}$ in def2-TZVP and 0.08 m$E_\mathrm{h}$ in def2-QZVP.

\begin{figure*}[p]
\centering
\includegraphics[width=0.85\textwidth]{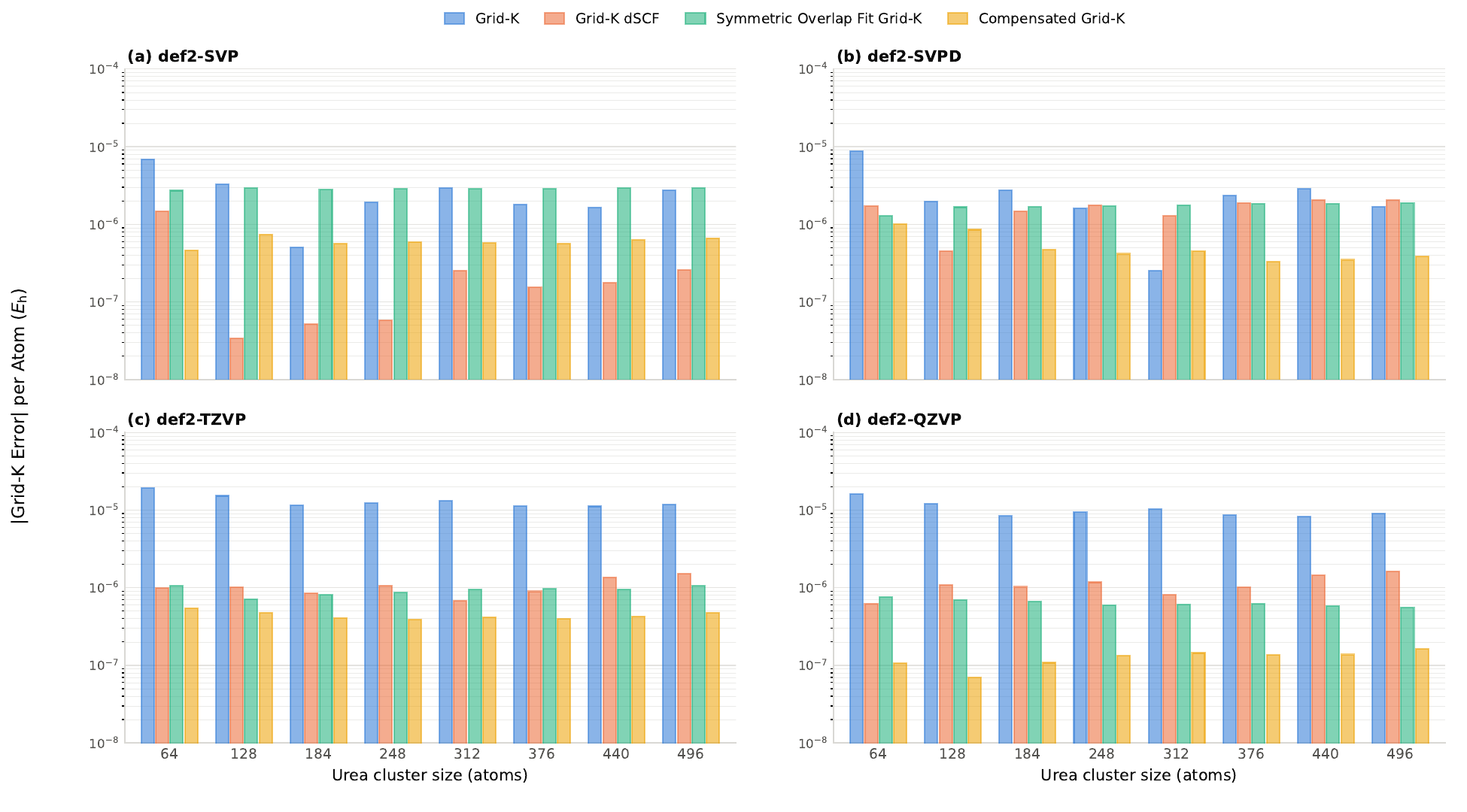}
\caption{Magnitude of the signed error per atom in the $\omega$B97M-V total
energy, $|E^{\mathrm{Grid\text{-}K}} - E^{\mathrm{analytic}}| /
N_{\mathrm{atom}}$, on a logarithmic scale, for clusters of 8 to 62 urea
molecules (64 to 496 atoms) cut from the urea crystal structure, computed
with canonical Grid-K, Grid-K with the SAD correction (dSCF), the symmetric
orthogonal overlap fit, and compensated Grid-K on the SG-0 exchange grid in
the (a) def2-SVP, (b) def2-SVPD, (c) def2-TZVP, and (d) def2-QZVP basis sets.
The errors are measured against total energies computed with analytic
two-electron integrals at the same functional and basis set.}
\label{fig:abs_error_per_atom}
\end{figure*}

Since the exchange energy is a size extensive quantity, we expect the error
due to the Grid-K approximation to grow linearly with system size.
Dividing by the number of atoms, Figure \ref{fig:abs_error_per_atom}, makes
this size dependence explicit. Beyond the smallest clusters the per-atom error
of each variant is essentially constant, so the grid error is extensive and a
single per-atom figure characterizes each variant and basis set. Canonical
Grid-K carries up to 9 $\mu E_\mathrm{h}$ per atom in the double-zeta sets,
with a sign that changes between clusters in def2-SVPD, and 8--19
$\mu E_\mathrm{h}$ per atom in def2-TZVP and def2-QZVP. The symmetric
orthogonal overlap fit alone gives 2.9, 1.8, 1.0, and 0.6 $\mu E_\mathrm{h}$
per atom in def2-SVP, def2-SVPD, def2-TZVP, and def2-QZVP, decreasing
steadily with basis-set quality, while the SAD correction alone fluctuates
between 0.03 and 2 $\mu E_\mathrm{h}$ per atom with no systematic value,
which is why it cancels less reliably in relative energies than the
overlap-fitted variants. Compensated Grid-K settles at 0.6
$\mu E_\mathrm{h}$ per atom in def2-SVP, 0.4 in def2-SVPD and def2-TZVP, and
0.13 in def2-QZVP, with
a consistent sign in each basis set, so the absolute error of a compensated
Grid-K total energy can be anticipated from the atom count alone for a
particular class of molecule and improves with basis-set quality.

\subsection{Relative Property Accuracy}

\begin{figure*}[p]
\centering
\includegraphics[width=0.85\textwidth]{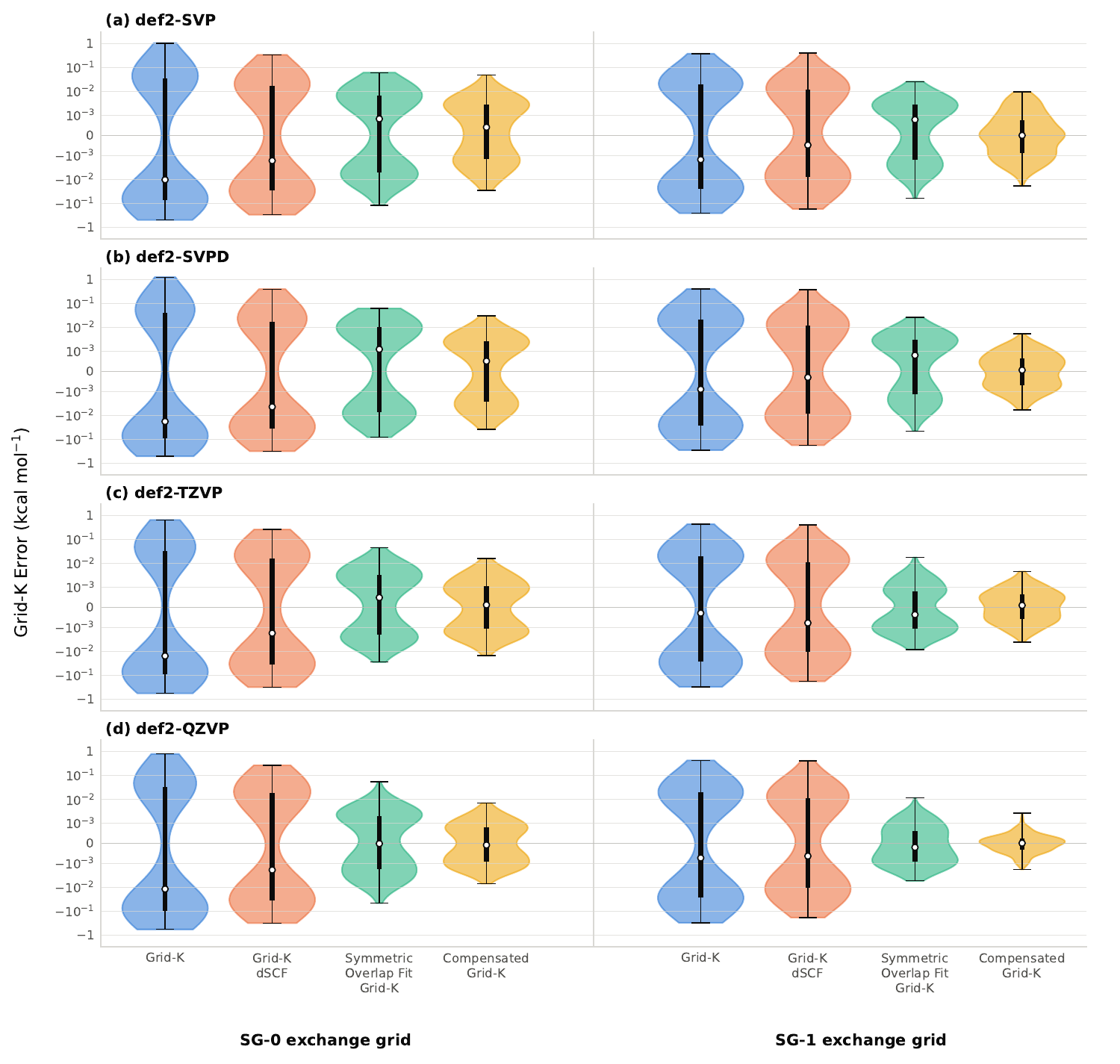}
\caption{Signed errors in $\omega$B97M-V relative energies computed with Grid-K variants with respect to analytic two-electron integrals for the 16 subsets of the GMTKN55 database.\cite{Goerigk:2017:32184} Each distribution pools the 340 benchmark reactions of all 16 subsets. Rows (a)--(d) are the def2-SVP, def2-SVPD, def2-TZVP, and def2-QZVP basis sets; within each row, the four Grid-K variants are shown for the SG-0 exchange grid (left group) and the SG-1 exchange grid (right group). Violins are colored by variant: canonical Grid-K (blue), Grid-K with the dSCF correction (orange), Grid-K with the symmetric orthogonal overlap fit (green), and compensated Grid-K, which combines the overlap fit with the dSCF correction (yellow). White dots mark medians, thick bars the interquartile ranges, and thin whiskers the full range. The vertical axis (Grid-K error) is arcsinh-scaled: linear within $\pm 3\times 10^{-4}$ kcal mol$^{-1}$ of zero and logarithmic beyond.}
\label{fig:gmtkn55_wb97mv}
\end{figure*}

\begin{figure*}[p]
\centering
\includegraphics[width=0.85\textwidth]{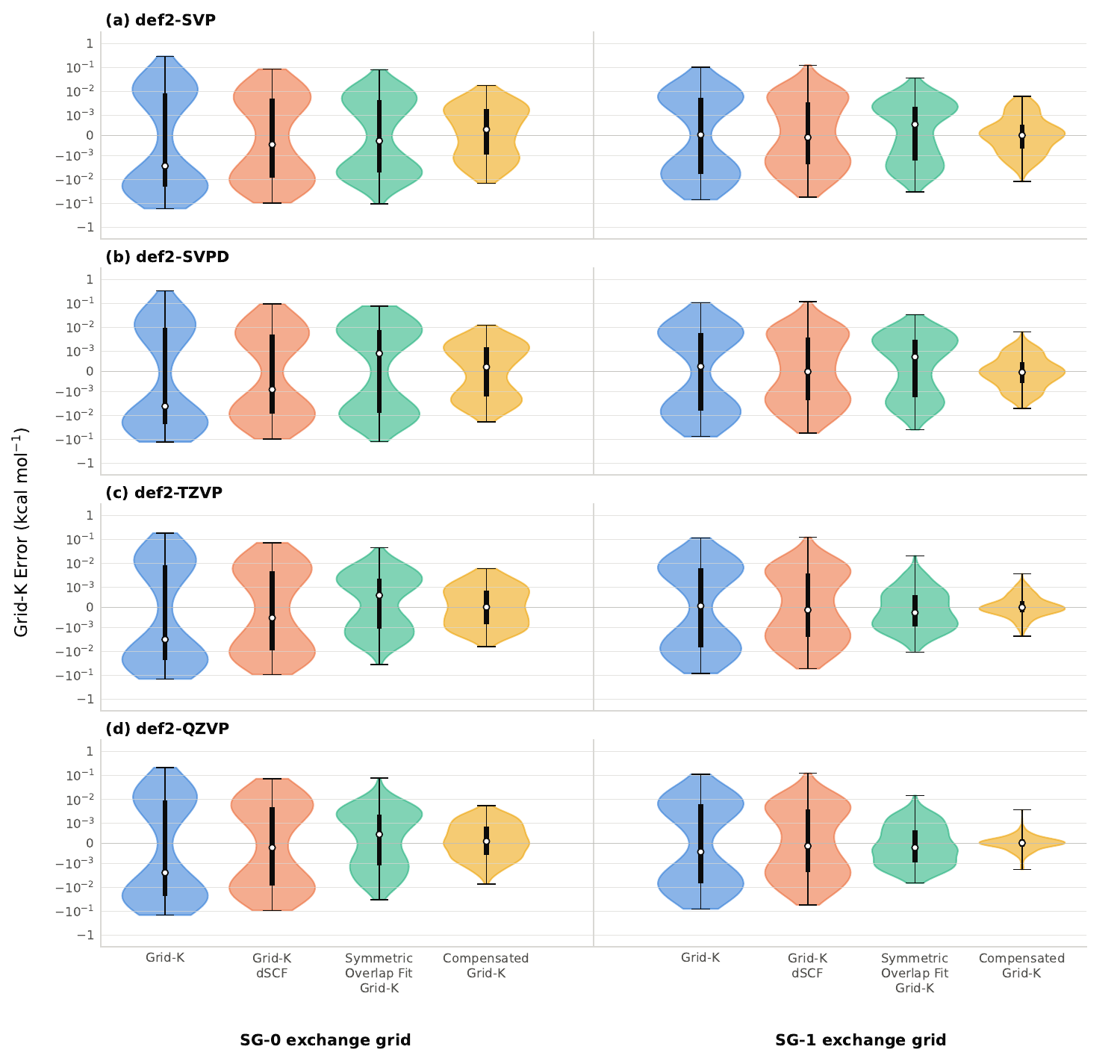}
\caption{Signed errors in B3LYP relative energies computed with Grid-K variants with respect to analytic two-electron integrals for the 16 subsets of the GMTKN55 database.\cite{Goerigk:2017:32184} Each distribution pools the 340 benchmark reactions of all 16 subsets. Rows (a)--(d) are the def2-SVP, def2-SVPD, def2-TZVP, and def2-QZVP basis sets; within each row, the four Grid-K variants are shown for the SG-0 exchange grid (left group) and the SG-1 exchange grid (right group). Violins are colored by variant: canonical Grid-K (blue), Grid-K with the dSCF correction (orange), Grid-K with the symmetric orthogonal overlap fit (green), and compensated Grid-K, which combines the overlap fit with the dSCF correction (yellow). White dots mark medians, thick bars the interquartile ranges, and thin whiskers the full range. The vertical axis (Grid-K error) is arcsinh-scaled: linear within $\pm 3\times 10^{-4}$ kcal mol$^{-1}$ of zero and logarithmic beyond.}
\label{fig:gmtkn55_b3lyp}
\end{figure*}

The performance of compensated Grid-K for relative energies is assessed over
16 subsets of the GMTKN55 database\cite{Goerigk:2017:32184} and summarized in
Figure \ref{fig:gmtkn55_wb97mv} for $\omega$B97M-V and Figure
\ref{fig:gmtkn55_b3lyp} for B3LYP, where each distribution pools all 340
benchmark reactions; the underlying total energies of every structure,
organized by class and subset, are provided in the Supporting Information.
The subsets span four classes of
relative energies: barrier heights of diverse main-group reactions (BHDIV10),
pericyclic reactions (BHPERI), single-bond rotations (BHROT27), and proton
exchange in small hydrogen-bonded clusters (PX13); conformational energies of
alkanes (ACONF), inorganic systems (ICONF), peptides (PCONF21), and sugars
(SCONF); reaction energies of double-bond isomerizations (CDIE20),
Diels--Alder reactions (DARC), organic isomerizations (ISO34), and protomer
equilibria (PArel); and intermolecular interaction energies of anionic
hydrogen-bonded dimers (AHB21), cation complexes (CHB6), ionic-liquid ion
pairs (IL16), and neutral noncovalent dimers (S66), for a total of 340
benchmark relative energies. Each relative
energy was evaluated with $\omega$B97M-V and B3LYP in the def2-SVP, def2-SVPD,
def2-TZVP, and def2-QZVP basis sets (2720 relative energies per method variant),
and the errors of the Grid-K variants are measured against relative energies
computed with analytic two-electron integrals at the same functional and basis
set. The same trends appear in all four property classes, and the three
ingredients of compensated Grid-K contribute nearly independently: the SAD correction reduces the errors of canonical Grid-K
by a factor of 2--3, the symmetric orthogonal overlap fit by roughly an order
of magnitude, and moving from the SG-0 to the SG-1 exchange grid by a further
factor of 2--3. Even on the coarse SG-0 grid, compensated Grid-K reproduces the
analytic-integral relative energies with a mean absolute error of 0.002
kcal mol$^{-1}$ and a worst case of 0.05 kcal mol$^{-1}$ over all 2720 data
points; on SG-1 the mean absolute error drops to 0.0007 kcal mol$^{-1}$ with no
error exceeding 0.02 kcal mol$^{-1}$. The largest errors of the uncorrected
canonical method occur for the conformer and reaction-energy classes (e.g.,
a maximum error of 1.24 kcal mol$^{-1}$ for an ICONF conformer with SG-0),
precisely the cases where error cancellation between similar structures is
least effective, and it is there that the compensation scheme yields its
largest improvements.

To probe the accuracy of compensated Grid-K for larger systems, we computed
protein--ligand interaction energies for the MMP12 target of the PL-REX
data set.\cite{Pecina:2024:1127} The 18 complexes pair an active-site cutout
of human matrix metallopeptidase 12 (728 atoms, including the catalytic
Zn$^{2+}$, total charge $+1$) with neutral inhibitors of 25--57 atoms, giving
complexes of 753--785 atoms; interaction energies are computed with monomers
frozen at their in-complex geometries and without counterpoise correction.
Table \ref{tab:plrex_mmp12} lists the $\omega$B97M-V interaction energies
obtained with compensated Grid-K on the SG-0 exchange grid, evaluating the
Coulomb term either analytically (direct J) or by density fitting (DF-J), in
the def2-SVP, def2-TZVP, and def2-TZVPP basis sets (up to nearly 18000
basis functions). Where fully analytic reference calculations are available,
the errors are shown in parentheses. With direct J, compensated Grid-K
reproduces the analytic interaction energies -- which span $-65$ to $-190$
kcal mol$^{-1}$ -- to within 0.025 kcal mol$^{-1}$ in def2-SVP and within
0.008 kcal mol$^{-1}$ in def2-TZVP, mirroring the basis-set trend observed
for GMTKN55 and demonstrating that the grid errors do not grow to chemical
significance even for systems approaching 800 atoms. The DF-J variant adds a
small, nearly uniform shift of $+0.02$ to $+0.08$ kcal mol$^{-1}$ that is
attributable to the density fitting of the Coulomb term and self-interaction
errors arising from differing treatments of Coulomb and exchange interactions.

\begin{figure*}[p]
\centering
\includegraphics[width=0.85\textwidth]{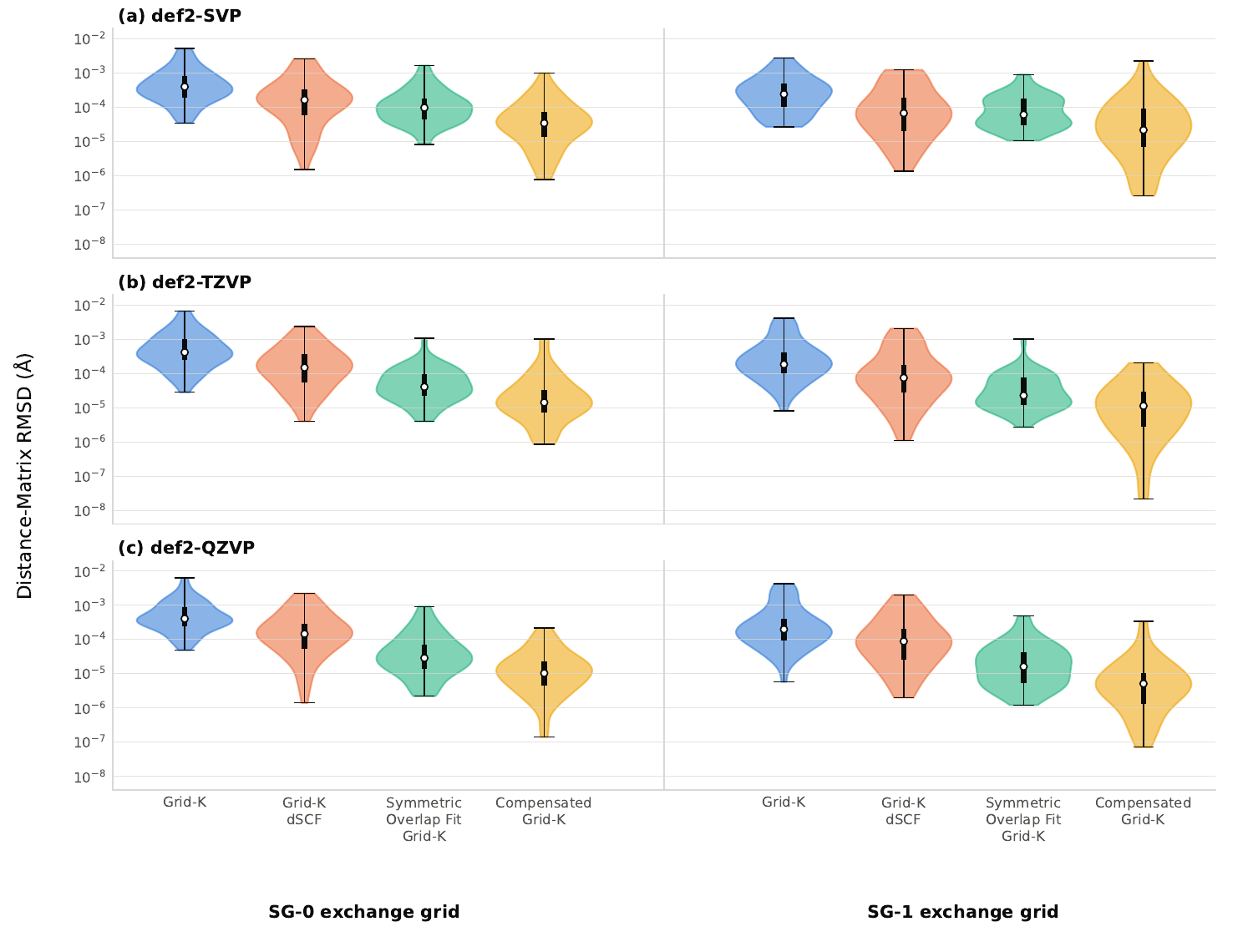}
\caption{RMSD between the interatomic distance matrices (over all unique atom
pairs) of geometries optimized with Grid-K variants and with analytic
two-electron integrals, for the 47 molecules of the B3se47
set\cite{Bremond:2016:459} at the $\omega$B97M-V level. Rows (a)--(c) are the def2-SVP, def2-TZVP, and
def2-QZVP basis sets; within each row, the four Grid-K variants are shown for
the SG-0 exchange grid (left group) and the SG-1 exchange grid (right group).
Violins are colored by variant: canonical Grid-K (blue), Grid-K with the dSCF
correction (orange), Grid-K with the symmetric orthogonal overlap fit
(green), and compensated Grid-K, which combines the overlap fit with the dSCF
correction (yellow). White dots mark medians, thick bars the interquartile
ranges, and thin whiskers the full range; the vertical axis is logarithmic.}
\label{fig:geom_rmsd}
\end{figure*}

Beyond single-point energies, we assessed the effect of Grid-K on optimized
geometries. The 47 molecules of the B3se47 set\cite{Bremond:2016:459} --
small and medium-sized organic molecules of 3--14 atoms containing H, C, N,
O, F, S, and Cl, including three cations -- were optimized with
$\omega$B97M-V in the def2-SVP,
def2-TZVP, and def2-QZVP basis sets using each Grid-K variant on the SG-0 and
SG-1 exchange grids as well as with analytic two-electron integrals, with a
tight gradient convergence threshold of $1.5\times10^{-5}$ Hartree/Bohr. To
compare geometries without rigid-body alignment, each optimized structure is
reduced to its matrix of interatomic distances, and the RMSD over the
$n(n-1)/2$ unique atom pairs between the Grid-K and analytic-integral optimized
geometries is computed. The distributions of these distance-matrix RMSDs are
shown in Figure \ref{fig:geom_rmsd}. Even canonical Grid-K on the SG-0 grid
perturbs the optimized geometries only mildly (median RMSD of
$4\times10^{-4}$ \AA{}, worst case $7\times10^{-3}$ \AA{}), and each
ingredient of the compensation scheme tightens the distributions further.
With compensated Grid-K the median distance-matrix RMSD is reduced by one to
two orders of magnitude relative to the canonical method, reaching
$\sim\!10^{-5}$ \AA{} with worst cases of $1$--$2\times10^{-3}$ \AA{} in
def2-SVP and def2-TZVP and $\sim\!3\times10^{-4}$ \AA{} in def2-QZVP --
far below the uncertainties relevant to any structural application, and, as
for the energies, improving systematically with basis-set quality.

\subsection{Analytic Gradient Check}

To verify that the Grid-K gradients derived above are the exact derivatives
of the Grid-K energies, we compare the analytic gradients computed with cuEST
against finite differences for methylpropyl isocyanate at the
$\omega$B97M-V/def2-QZVP level of theory. We make this comparison with grids
spanning roughly a factor of 50x difference in the number of points per atom.
We start with an extremely coarse SG-0(-4) grid (where derivatives of the
grid point motion and Becke weights contribute significantly to the gradient)
and increase to a much denser SG-3 grid. First, we consider the gradient of
the SCF energy using the canonical Grid-K approach for exchange. We find
agreement between the analytic derivative and finite difference derivative
to better than $10^{-9}$ $E_\mathrm{h}$ bohr$^{-1}$ in the largest component
of the gradient. When the symmetric overlap fit is used, the error in the
$\omega$B97M-V gradient remains effectively unchanged. Next, we isolated the 
exchange gradient contribution. The derivative of the exchange energy is 
evaluated using finite difference with a fixed density matrix obtained from
the SCF at the central geometry. Since the exchange gradient does not 
have noise from SCF convergence, the accuracy in this term is slightly 
better with agreement to better than $10^{-10}$ $E_\mathrm{h}$ bohr$^{-1}$
in all cases. We note that the accuracy of the canonical Grid-K gradient
is slightly better; this can likely be attributed to noise originating from
the eigensolver that is introduced while forming the transfer matrix.
This extremely tight agreement between the finite difference and analytic
derivatives indicates that the formulation (and implementation) of the 
Grid-K gradients is correct both with and without application of the
symmetric overlap fitting. Importantly, the SCF gradients and Grid-K 
gradients show no dependence whatsoever on the grid 
quality: the agreement is the same on the coarsest SG-0(-4) grid as on SG-3,
and the same with and without the overlap fit.

In passing, we revisit the temptation mentioned earlier to approximate the analytic 
gradient using expressions obtained in the limit of a complete grid. In this 
limit, only the basis function advection of
the collocation matrix is retained (and doubled to stand in for the four symmetric
contributions of Eq.~\eqref{eq:hybrid-exchange-nuclear-gradient}).
We find extremely poor performance of this approximation using coarse grids as
is expected and we find that the approximation degrades further when symmetric
overlap fitting is applied.
Even on SG-3 the complete grid limit (CGL) deviation exceeds the agreement between
the analytic gradient and the finite difference by at least five orders of
magnitude and remains larger than the gradient convergence threshold of the
geometry optimizations in the previous subsection, so the heuristic is not a
substitute for the analytic gradient on any grid of practical size.

\subsection{Computational Efficiency}

\begin{figure*}[p]
\centering
\includegraphics[width=0.72\textwidth]{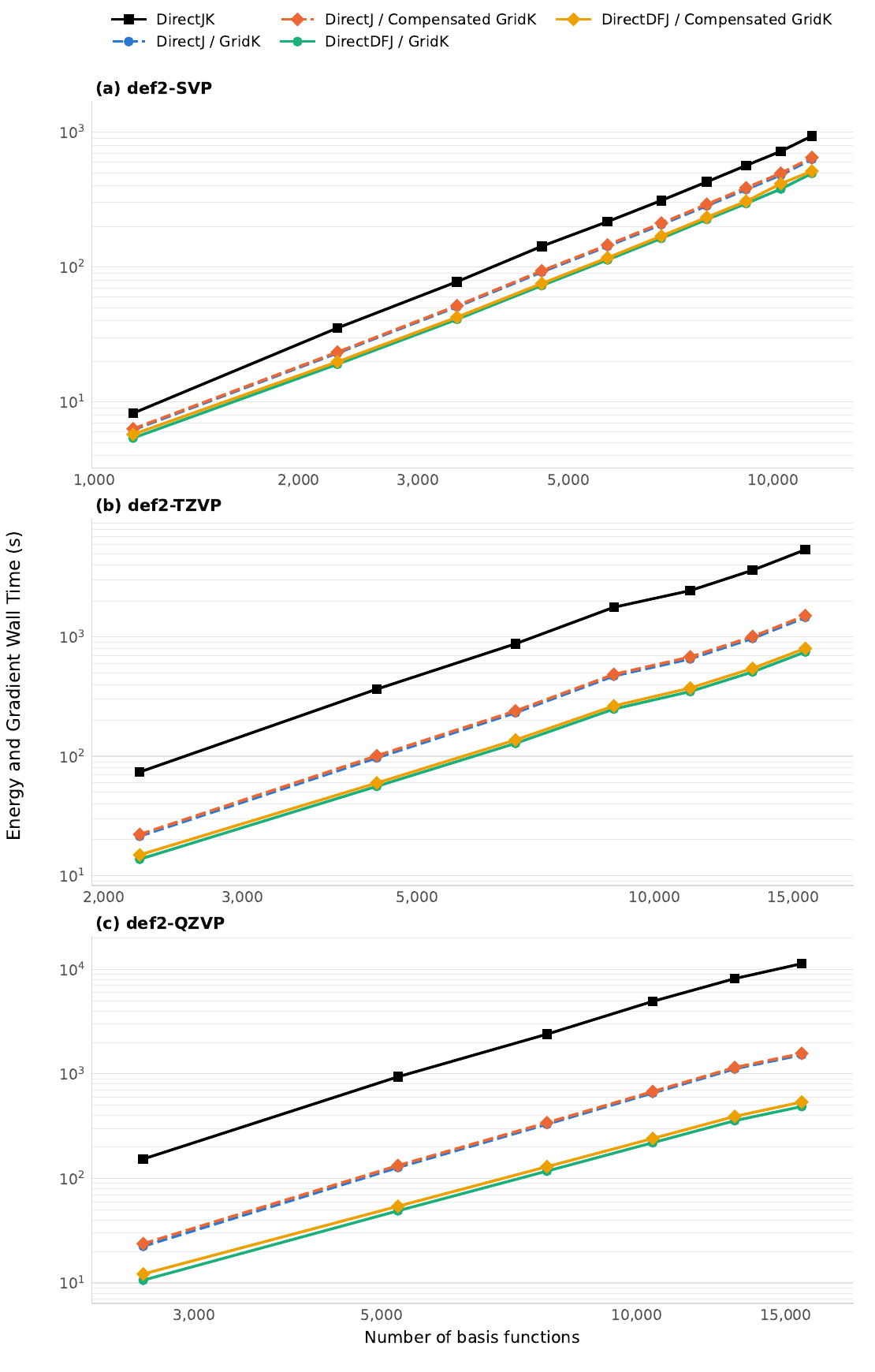}
\caption{Wall time for a B3LYP energy and analytic gradient on a single NVIDIA
B200 GPU for clusters of 5 to 100 molecules cut from the benzene crystal
structure, as a function of the number of basis functions, in the (a)
def2-SVP, (b) def2-TZVP, and (c) def2-QZVP basis sets. Integral-direct J and K (DirectJK) is
compared with canonical Grid-K and compensated Grid-K on the SG-0 exchange
grid, each paired with either integral-direct J (DirectJ) or density-fitted J
(DF-J). Both axes are logarithmic.}
\label{fig:timings}
\end{figure*}

To quantify the cost of Grid-K relative to integral-direct exchange, we
computed B3LYP energies and analytic gradients for a sequence of clusters cut
from the benzene crystal structure, containing 10 to 100 molecules (120 to
1200 atoms, up to 11400 basis functions) in the def2-SVP basis set, 10 to
70 molecules (up to 15540 basis functions) in the def2-TZVP basis set, and 5
to 30 molecules (up to 15660 basis functions) in the def2-QZVP basis set, on
a single NVIDIA B200 GPU. The exchange grid is SG-0 throughout. Five variants
are compared: integral-direct J and K (DirectJK), and Grid-K or compensated
Grid-K paired with either integral-direct J (DirectJ) or density-fitted J
(DF-J). For these timings, DF-J uses the def2-universal-JFIT auxiliary
basis.\cite{Weigend:2006:1057} Figure \ref{fig:timings} shows the total wall
time for the energy and
gradient as a function of the number of basis functions. All variants
converge in 12 to 14 SCF iterations, with iteration counts that agree across
methods at every cluster size but one, where they differ by a single
iteration, so the timings reflect the cost of the Fock and
gradient builds, with the gradient accounting for roughly one fifth to one
quarter of the total in every case. All curves scale roughly quadratically
with the basis-set size in this regime, so the Grid-K variants preserve the
asymptotic behavior of the integral-direct reference and shift it by a
constant factor that grows rapidly with the basis-set cardinality. The
additional cost of the compensation scheme is modest: with integral-direct J,
the overlap fit and SAD exchange terms add 2--6\% to the wall time of
canonical Grid-K, and with DF-J, where the exchange terms make up a larger
share of the total, 4--9\% in def2-SVP and def2-TZVP and 10--15\% in
def2-QZVP.

\begin{figure*}[p]
\centering
\includegraphics[width=0.72\textwidth]{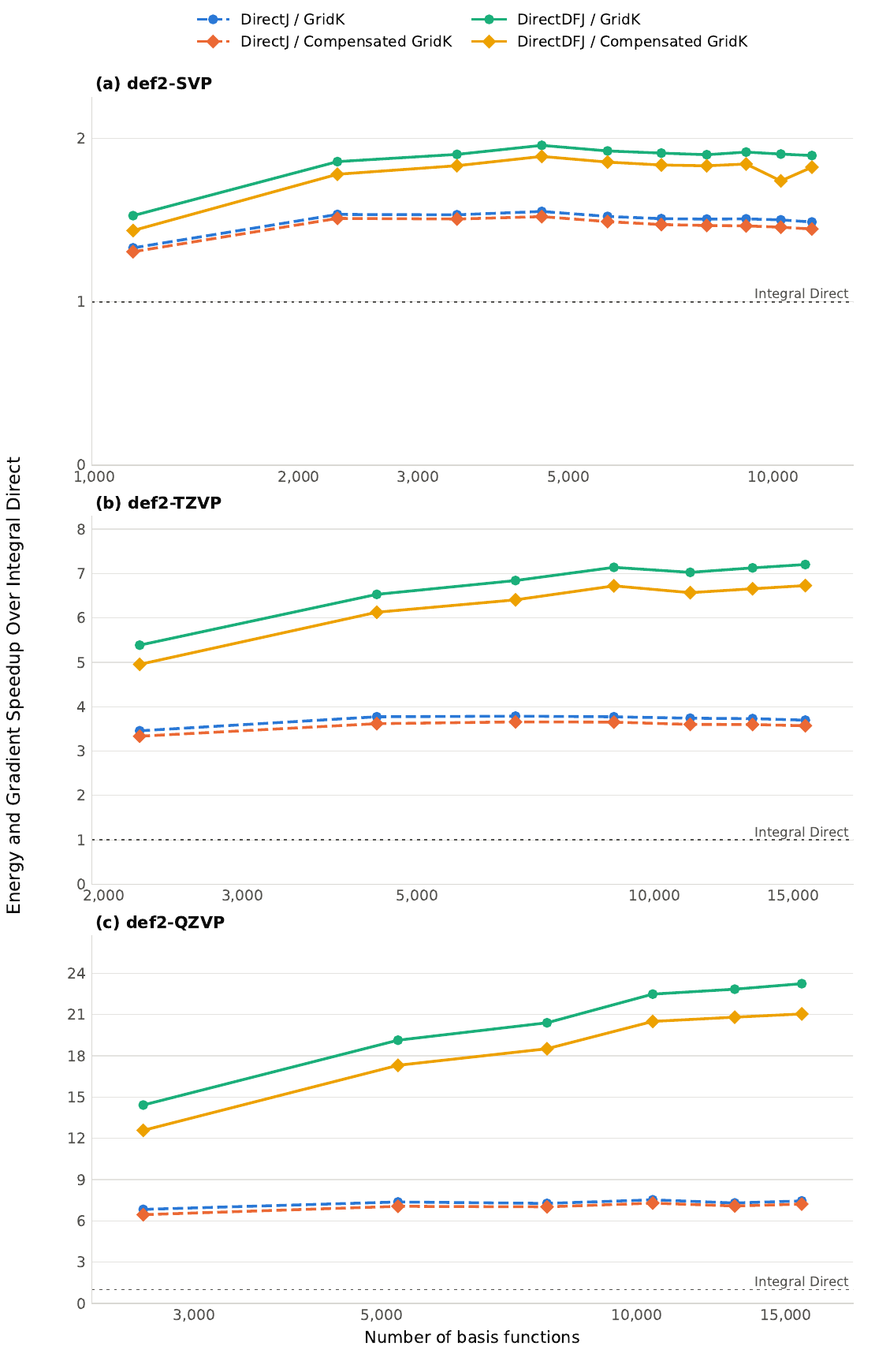}
\caption{Speedup of the B3LYP energy and analytic gradient wall time over
integral-direct J and K for the benzene crystal clusters of Figure
\ref{fig:timings}, in the (a) def2-SVP, (b) def2-TZVP, and (c) def2-QZVP
basis sets, for canonical Grid-K and compensated Grid-K on the SG-0 exchange grid paired with
either integral-direct J (DirectJ) or density-fitted J (DF-J). The dotted line
marks the integral-direct reference.}
\label{fig:speedup}
\end{figure*}

Figure \ref{fig:speedup} shows the same data as speedups over DirectJK. The
speedup rises over the smallest clusters, where fixed overheads are
proportionally larger, and plateaus beyond a few thousand basis functions.
In def2-SVP, Grid-K with integral-direct J is about 1.5$\times$ faster than
DirectJK, and replacing the Coulomb build with DF-J raises this to about
1.9$\times$. In def2-TZVP the gains are substantially larger, 3.6--3.8$\times$
with integral-direct J and 6.7--7.2$\times$ with DF-J, and in def2-QZVP they
reach 7.0--7.5$\times$ with integral-direct J and 19--23$\times$ with DF-J.
The speedup thus roughly doubles with each step in basis-set cardinality
when the Coulomb build is integral-direct, and roughly triples when it is
density fitted, as the high-angular-momentum functions that dominate the cost
of the analytic ERIs are handled by collocation in Grid-K. Compensated Grid-K
retains most of the speedup of canonical Grid-K, with the gap widening from a
few percent in def2-SVP to about 10\% in def2-QZVP with DF-J, while delivering
the accuracy improvements of the preceding sections.

\begin{figure*}[p]
\centering
\includegraphics[width=\textwidth]{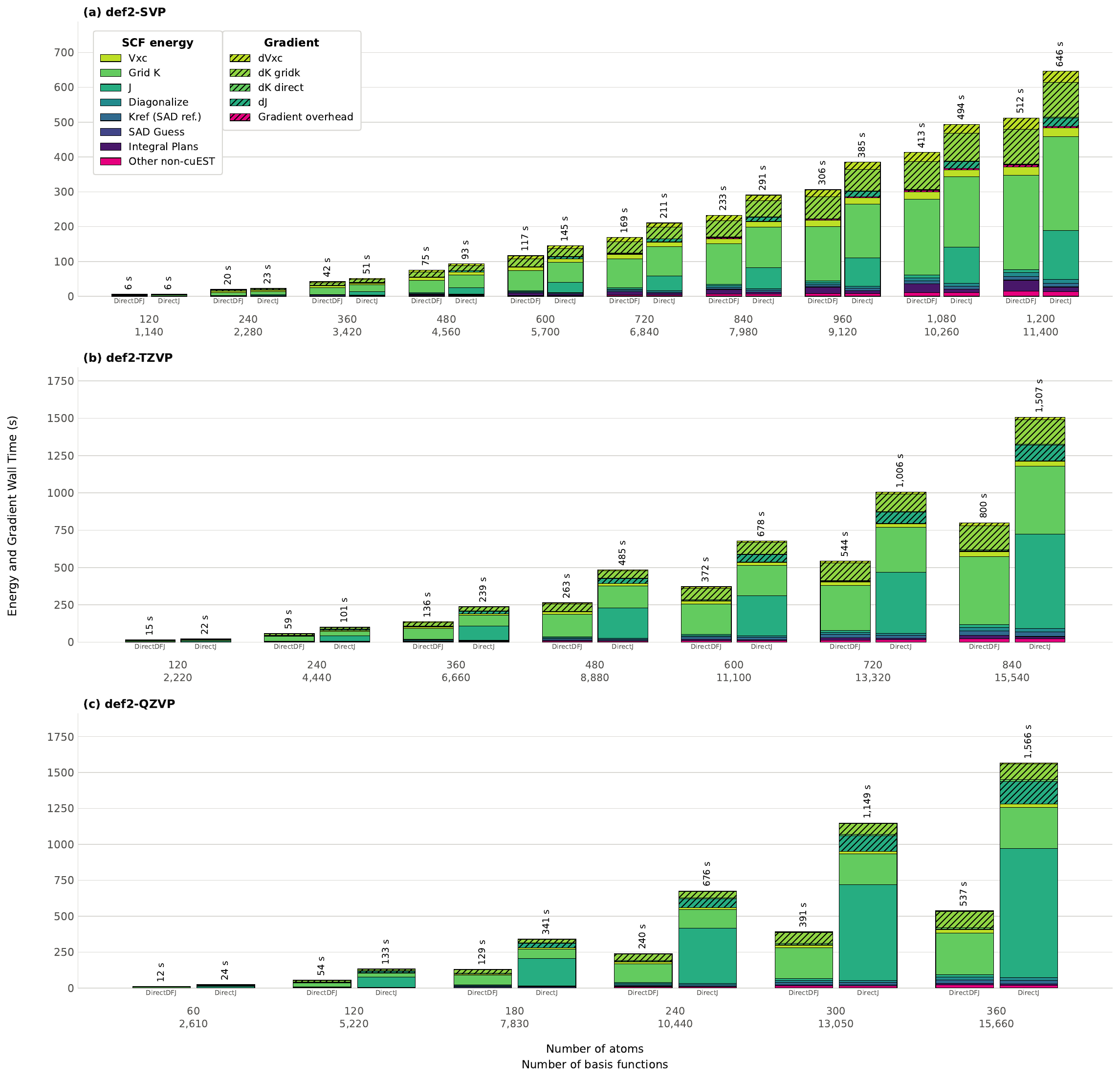}
\caption{Decomposition of the compensated Grid-K B3LYP energy and analytic
gradient wall time on a single NVIDIA B200 GPU for the benzene crystal
clusters of Figure \ref{fig:timings}, in the (a) def2-SVP, (b) def2-TZVP, and
(c) def2-QZVP basis sets. At each cluster size the left bar uses density-fitted J (DF-J)
and the right bar integral-direct J (DirectJ), both with compensated Grid-K
on the SG-0 exchange grid. Solid bands are SCF energy terms and hatched bands
are gradient terms, each gradient band taking the color of the energy term it
differentiates. Kref is the one-time SAD reference exchange matrix and dK
direct the SAD bilinear gradient term; both are specific to the compensation
scheme.}
\label{fig:breakdown}
\end{figure*}

Figure \ref{fig:breakdown} decomposes the compensated Grid-K wall times into
their components, with the SCF energy terms as solid bands and the gradient
terms as hatched bands, for the DF-J and integral-direct J variants side by
side at each cluster size. With integral-direct J, the Coulomb matrix and its
gradient account for roughly one quarter of the total in def2-SVP, one half
in def2-TZVP, and two thirds in def2-QZVP, so the Coulomb build, rather than
exchange, is the largest single cost in the larger basis sets once Grid-K is
used. Replacing it with DF-J reduces the Coulomb share to a few percent,
after which the Grid-K exchange matrix and its gradient dominate, accounting
for roughly three quarters of the total in all three basis sets, with the
exchange--correlation quadrature, diagonalization, integral plans, and
non-cuEST overhead making up the remainder. The terms specific to the
compensation scheme remain small: the one-time SAD reference exchange matrix
and the SAD bilinear gradient together account for 2--5\% of the total in
def2-SVP and def2-TZVP and 7--11\% in def2-QZVP, where the analytic SAD
exchange matrix inherits the steep basis-set dependence of the ERIs, and the
overlap-fitting linear algebra is not visible at this scale.

\section{Summary and Outlook}

This paper introduces the compensated Grid-K scheme for hybrid exchange and
first derivatives thereof. Augmenting the plain Grid-K seminumerical hybrid
exchange scheme with dSCF difference density and symmetric overlap fitting
(orthogonal orbital basis) compensations is found to yield a differentiable
method with dramatically reduced absolute and relative error properties, while
retaining most of the conceptual simplicity of the original Grid-K approach. The
approach is well suited to implementation on modern GPUs, and we provide an
efficient implementation of the major components of compensated Grid-K within
the recently released cuEST v0.3 CUDA-X library for quantum chemistry. 

We particularly recommend the compensated Grid-K method for use on large-scale
molecular systems in high-quality def2-TZVP or def2-QZVP basis sets. These basis
sets are large enough that the symmetric overlap-fitting procedure provides large
accuracy improvements when used together with high-quality MultiExp
based grids like SG-0, and even trimmed versions like SG-0($-2$). Some quick
figures of merit from the tour section: Gramicidin dimer end-to-end SCF time in
def2-TZVP (10112 nao) is 294 s, in def2-QZVP (23904 nao) is 1259 s, all in
$\omega$B97M-V, SG-0($-2$) compensated Grid-K and DF-J, with $<0.1$ kcal
mol$^{-1}$ errors in interaction energies dominated by the widely trusted DF-J
component. This is extremely high throughput for this basis quality, with K
components accelerated over DirectK by $16\times$ and $57\times$ in def2-TZVP
and def2-QZVP, respectively.

One of the strengths of compensated Grid-K is the existence of an analytical
derivative of the approximate hybrid exchange energy including all basis
function, Becke grid, and compensation contributions -- this is derived herein
and implemented component by component in cuEST v0.3. The gradient has a
particularly elegant Sylvester equation form that is easy to implement.  

A few notes on reservations and cautions. One is that while the general field of
seminumerical exchange is over 40 years old, the specific recipe of compensated
Grid-K is brand new and deserves some additional benchmarking. In particular,
while our early benchmarks show that augmented basis sets like def2-SVPD perform
well with SG-0 and SG-1, it is our experience that grid-based methods with
augmented functions almost always require customized grids for best performance,
particularly in high zeta and/or multiple augmentation cases. Indeed, a
thorough grid optimization and benchmarking study is merited for Grid-K -- this
might yield reliable grids of SG-0($-2$) size that provide robust
performance for many absolute and relative observable properties. Another
caution is def2-SVP (and smaller non-augmented basis sets) -- the smaller span
of this basis set does not benefit as much as higher zeta basis sets from the
symmetric overlap-fitting procedure, and incurs slightly higher observable
errors than def2-TZVP or def2-QZVP. Colloquially we take this as a sign that
perhaps it is time to switch to efficient DirectJ/K techniques for def2-SVP and
smaller, which are anyway near to the computational crossover point with
compensated Grid-K. 

There are some straightforward extensions and some open questions around
compensated Grid-K. At a tactical level, hardening compensated Grid-K to feature
robust optimized grids, deeper derivatives, deployment in non-symmetric and
response environments (both possible with cuEST v0.3), and dSCF-corrected DF-J
are straightforward and will surely be pursued by the cuEST team and our
collaborators. Mixed-precision approaches to enhance applicability to FP64-lite
GPUs are surely worth pursuing, including some of our recent
float-float\cite{Dekker:1971:224,Thall:2006:52} and Ozaki
GEMM\cite{Ozaki:2012:95,Ozaki:2026:IJHPCA} acceleration techniques from work on
CoreDF.\cite{cuEST:2026:GTC} At an intermediate level, we point out that there
seems to exist an interesting tension in seminumerical Grid-K methods between
compensating and not compensating, particularly where overlap fitting is
concerned. If we choose to use overlap-fitting compensation, we can realize
extremely small errors, but generally must pay for this by working with formally
dense intermediates and performing formally $\mathcal{O}(N^3)$ linear algebraic
steps like dense eigensolves. If we choose not to use overlap-fitting
compensation, more advantage can be made of sparsity in the collocation, but
seemingly at the cost of very high error tradeoffs. For this instance of
compensated Grid-K we have made the deliberate trade of overlap-fitting
compensation in lieu of sparsity, and rely on the exceptional throughput of
GPUs, cuBLAS, and cuSOLVER to cover this. Longer term it is worth asking if one
can have one's cake and eat it on this topic, particularly where symmetric
overlap fitting in an orthogonal basis is concerned. At a more strategic level,
we will pose a question that has plagued some of us for more than a decade: why
does the pragmatically best method have one grid integral and one analytical
integral? Direct has both integrals analytical. THC has both integrals
grid-based. Compensated Grid-K has the extremely inelegant symmetry-broken
choice of one each and seems to provide a very interesting mix of speed and
accuracy. Note that this also happens sometimes in correlated methods like
PS/THC-MP2-K\cite{Martinez:1994:3631,Murphy:1995:1481} or the ppL term of
PS/THC-CCSD.\cite{Parrish:2014:181102}

Like the Last Alliance of Elves and Men, we quantum chemists have fought K for
decades. Today's result may provide a hard-won reprieve in that long campaign,
with large reductions for high-zeta K at negligible and differentiable
approximation levels. However, we caution that our enemy has not been entirely
defeated -- a quick look at the Gramicidin dimer def2-TZVP and def2-QZVP results
in the smallest SG-0($-2$) grid shows that compensated Grid-K remains just under
$60\%$ of the total SCF. K now lurks in the shadows, and there will doubtless
come a day soon when we must face this familiar foe again. 

\textbf{AI Disclosure:} Agentic AI was used to accelerate human effort in many
stages of this paper, including method validation, code generation, benchmark
orchestration, analysis, paper writing, and figure script generation. All
numerical data and figures come from numerical experiments obtained using cuEST.
The human authors certify all content and take responsibility for the scientific
correctness of the work.

\textbf{Conflict of Interest:} All authors are employees of NVIDIA Corporation,
which develops the cuEST CUDA-X library and the GPUs used in this work, and hold
NVIDIA stock and/or stock options.

\textbf{Data Availability:} The raw data supporting the findings of this
study, including structures, output log files, and example scripts to
reproduce the results, are available as Supporting Information in the
journal form of this article. The calculations were performed with CTK 13.4
and a pre-release version of cuEST v0.3; cuEST is available for download at
\url{https://developer.nvidia.com/cuest-downloads}.

\bibliography{jrncodes,refs}

\begin{table*}[p]
\caption{Protein--ligand interaction energies (kcal mol$^{-1}$) for the 18
MMP12 complexes of the PL-REX data set,\cite{Pecina:2024:1127} computed with
$\omega$B97M-V using compensated Grid-K on the SG-0 exchange grid with the
Coulomb term evaluated analytically (direct J) or by density fitting (DF-J).
Monomers are frozen at their in-complex geometries and no counterpoise
correction is applied. Errors with respect to fully analytic two-electron
integrals are given in parentheses where the reference calculations are
available; the error statistics are evaluated over all 18 complexes.}
\label{tab:plrex_mmp12}
\begin{ruledtabular}
\begin{tabular}{lcccccc}
 & \multicolumn{2}{c}{def2-SVP} & \multicolumn{2}{c}{def2-TZVP} & \multicolumn{2}{c}{def2-TZVPP} \\
\cline{2-3}\cline{4-5}\cline{6-7}
Complex & direct J & DF-J & direct J & DF-J & direct J & DF-J \\
\hline
3EHX & $-155.32$ ($-0.012$) & $-155.29$ ($+0.023$) & $-131.38$ ($-0.002$) & $-131.31$ ($+0.068$) & $-130.74$ ($-0.002$) & $-130.68$ ($+0.052$) \\
3EHY & $-136.37$ ($-0.005$) & $-136.34$ ($+0.021$) & $-114.43$ ($+0.006$) & $-114.37$ ($+0.060$) & $-113.96$ & $-113.92$ \\
3F15 & $-163.26$ ($+0.025$) & $-163.23$ ($+0.055$) & $-136.59$ ($+0.004$) & $-136.52$ ($+0.076$) & $-136.14$ & $-136.09$ \\
3F16 & $-156.95$ ($+0.015$) & $-156.92$ ($+0.042$) & $-131.55$ ($+0.002$) & $-131.49$ ($+0.057$) & $-131.06$ & $-131.01$ \\
3F17 & $-158.46$ ($+0.010$) & $-158.43$ ($+0.040$) & $-133.99$ ($-0.001$) & $-133.92$ ($+0.065$) & $-133.36$ & $-133.31$ \\
3F18 & $-160.69$ ($+0.019$) & $-160.66$ ($+0.045$) & $-134.53$ ($+0.002$) & $-134.48$ ($+0.055$) & $-134.12$ & $-134.08$ \\
3F19 & $-144.59$ ($+0.012$) & $-144.57$ ($+0.036$) & $-120.79$ ($+0.002$) & $-120.74$ ($+0.046$) & $-120.42$ & $-120.38$ \\
3F1A & $-140.84$ ($+0.011$) & $-140.82$ ($+0.034$) & $-119.22$ ($+0.001$) & $-119.17$ ($+0.048$) & $-118.82$ & $-118.78$ \\
3LK8 & $-150.15$ ($+0.018$) & $-150.12$ ($+0.044$) & $-125.81$ ($+0.003$) & $-125.75$ ($+0.060$) & $-125.38$ & $-125.34$ \\
3N2U & $-189.98$ ($+0.005$) & $-189.94$ ($+0.040$) & $-157.40$ ($+0.000$) & $-157.32$ ($+0.079$) & $-156.70$ & $-156.64$ \\
3N2V & $-158.94$ ($+0.002$) & $-158.91$ ($+0.036$) & $-134.16$ ($+0.002$) & $-134.09$ ($+0.077$) & $-133.54$ & $-133.48$ \\
3NX7 & $-155.31$ ($+0.022$) & $-155.29$ ($+0.049$) & $-129.46$ ($+0.006$) & $-129.40$ ($+0.072$) & $-129.02$ & $-128.97$ \\
3RTS & $-152.35$ ($+0.012$) & $-152.32$ ($+0.038$) & $-128.10$ ($+0.005$) & $-128.05$ ($+0.057$) & $-127.62$ & $-127.58$ \\
3RTT & $-153.89$ ($+0.020$) & $-153.85$ ($+0.055$) & $-128.87$ ($+0.008$) & $-128.81$ ($+0.075$) & $-128.41$ & $-128.36$ \\
4GUY & $-165.36$ ($+0.005$) & $-165.33$ ($+0.038$) & $-138.17$ ($-0.003$) & $-138.09$ ($+0.075$) & $-137.53$ & $-137.47$ \\
5LAB & $-155.19$ ($+0.014$) & $-155.16$ ($+0.046$) & $-129.25$ ($+0.005$) & $-129.18$ ($+0.071$) & $-128.79$ & $-128.74$ \\
6RD0 & $-93.83$ ($-0.003$) & $-93.81$ ($+0.024$) & $-78.42$ ($-0.003$) & $-78.35$ ($+0.067$) & $-77.85$ & $-77.80$ \\
6RLY & $-80.13$ ($+0.009$) & $-80.11$ ($+0.034$) & $-64.94$ ($-0.007$) & $-64.87$ ($+0.064$) & $-64.26$ & $-64.21$ \\
\hline
Mean error & $+0.010$ & $+0.039$ & $+0.002$ & $+0.065$ & -- & -- \\
Std.\ deviation & $0.010$ & $0.010$ & $0.004$ & $0.010$ & -- & -- \\
Max abs.\ error & $0.025$ & $0.055$ & $0.008$ & $0.079$ & -- & -- \\
\end{tabular}
\end{ruledtabular}
\end{table*}

\begin{table*}[p]
\caption{Maximum absolute deviations ($E_\mathrm{h}$ bohr$^{-1}$) over the
$3 N_{\mathrm{atom}} = 48$ components of the gradient of methylpropyl
isocyanate at the $\omega$B97M-V/def2-QZVP level of theory, for Grid-K and Symmetric
Overlap Fit Grid-K on exchange grids from SG-0($-4$) to SG-3. For each
variant, ``SCF gradient fidelity'' compares the complete analytic gradient of the
$\omega$B97M-V energy with a four-point central finite difference of the
total SCF energy (step $3\times10^{-3}$ bohr); ``Exchange gradient fidelity'' compares 
$E_{K}^{\mathrm{grid},\xi}$ of Eq.~\eqref{eq:gridk_grad} or $\bar E_{K}^{\mathrm{grid}, \xi}$ of
Eq.~\eqref{eq:cgridk_grad_final} with a six-point central finite difference
of the Grid-K exchange energy (step $3\times10^{-4}$ bohr) with a fixed density matrix (the SCF density converged at the central geometry); 
and ``CGL gradient approximation'' is the deviation from $E_{K}^{\mathrm{grid},
\xi}$ or $\bar E_{K}^{\mathrm{grid}, \xi}$ of the complete-grid-limit (CGL)
approximation that retains only the doubled basis function advection of the
collocation matrix. Points per atom refer to this molecule, specifically.}
\label{tab:gradient_fd}
\begin{ruledtabular}
\begin{tabular}{lrcccccc}
 & & \multicolumn{3}{c}{Grid-K} & \multicolumn{3}{c}{Symmetric Overlap Fit Grid-K} \\
\cline{3-5}\cline{6-8}
Grid & Points/atom & \shortstack{SCF gradient\\fidelity} & \shortstack{Exchange gradient\\fidelity} & \shortstack{CGL gradient\\approximation} & \shortstack{SCF gradient\\fidelity} & \shortstack{Exchange gradient\\fidelity} & \shortstack{CGL gradient\\approximation} \\
\hline
SG-0($-4$) &   386 & $5.8\times10^{-10}$ & $1.5\times10^{-11}$ & $1.92\times10^{-2}$ & $5.5\times10^{-10}$ & $5.0\times10^{-11}$ & $8.00\times10^{-1}$ \\
SG-0($-3$) &   573 & $5.3\times10^{-10}$ & $1.6\times10^{-11}$ & $9.08\times10^{-3}$ & $5.7\times10^{-10}$ & $5.8\times10^{-11}$ & $2.02\times10^{-1}$ \\
SG-0($-2$) &   747 & $5.8\times10^{-10}$ & $1.1\times10^{-11}$ & $4.20\times10^{-3}$ & $6.4\times10^{-10}$ & $5.6\times10^{-11}$ & $1.89\times10^{-1}$ \\
SG-0($-1$) &  1010 & $4.4\times10^{-10}$ & $1.3\times10^{-11}$ & $1.21\times10^{-3}$ & $6.3\times10^{-10}$ & $5.5\times10^{-11}$ & $7.21\times10^{-2}$ \\
SG-0       &  1386 & $5.7\times10^{-10}$ & $1.3\times10^{-11}$ & $5.39\times10^{-4}$ & $3.2\times10^{-10}$ & $6.1\times10^{-11}$ & $3.16\times10^{-2}$ \\
SG-1       &  3780 & $9.1\times10^{-10}$ & $1.6\times10^{-11}$ & $2.68\times10^{-4}$ & $5.8\times10^{-10}$ & $5.4\times10^{-11}$ & $1.26\times10^{-2}$ \\
SG-2       &  7421 & $6.3\times10^{-10}$ & $1.0\times10^{-11}$ & $1.00\times10^{-4}$ & $4.9\times10^{-10}$ & $5.8\times10^{-11}$ & $2.68\times10^{-2}$ \\
SG-3       & 17250 & $4.7\times10^{-10}$ & $1.5\times10^{-11}$ & $7.80\times10^{-5}$ & $4.6\times10^{-10}$ & $5.1\times10^{-11}$ & $4.18\times10^{-3}$ \\
\end{tabular}
\end{ruledtabular}
\end{table*}

\end{document}